%% file: main.tex
\documentclass[lettersize,journal]{IEEEtran}
\usepackage{amsmath,amsfonts}
\usepackage[T1]{fontenc}
\usepackage{array}
\usepackage{textcomp}
\usepackage{stfloats}
\usepackage{url}
\usepackage{verbatim}
\usepackage{graphicx}
\usepackage{cite}
\usepackage{tikz}
\usepackage[acronym]{glossaries}
 
\include{acronyms.tex}

\usepackage{xcolor}
\usepackage{comment}
\usepackage{float}
\usepackage{subcaption}
\usepackage{dsfont}

\usepackage{mathtools}

\usepackage{algorithm}
\usepackage{algpseudocode}

\usepackage{wrapfig}
\usepackage{booktabs}
\usepackage{multirow}
\usepackage{physics}
\usepackage{bbold}
\usepackage[hidelinks]{hyperref}
\usepackage[shortlabels]{enumitem}

\newcommand{\rev}[1]{{\color{black}{#1}}}

\begin{document}
\title{Optimizing Entanglement Distribution Protocols:\\Maximizing Classical Information in Quantum Networks}

\author{Ethan Sanchez Hidalgo, Diego Zafra Bono, Guillermo Encinas Lago, J. Xavier Salvat Lozano,\\ Jose A. Ayala-Romero, Xavier Costa Perez,~\IEEEmembership{Senior Member,~IEEE}\vspace{-15pt}
\thanks{Ethan Sanchez Hidalgo and Diego Zafra Bono were with i2CAT Foundation, Barcelona, Spain (email: ethan.sanchez.hidalgo@gmail.com, diegozafrabono@gmail.com). Guillermo Encinas Lago is with i2CAT Foundation, Barcelona, Spain (e-mail: guillermo.encinas@i2cat.net). J. Xavier Salvat Lozano is with NEC Laboratories Europe GmbH, Heidelberg, Germany and Universitat Autonoma de Barcelona (UAB), Bellaterra, Spain (e-mail: josep.xavier.salvat@neclab.eu). Jose A. Ayala-Romero is with NEC Laboratories Europe GmbH, Heidelberg, Germany (e-mail: jose.ayala@neclab.eu). Xavier Costa-Pérez is with NEC Laboratories Europe GmbH, Heidelberg, Germany, i2CAT Foundation and ICREA, Barcelona, Spain (e-mail: xavier.costa@ieee.org).}
\thanks{Work supported by the European Commission through grants No. 101139270 (ORIGAMI) and SNS-JU-101192521 (Multi-X) and by the CERCA Programme.}}

\markboth{ }{}

\maketitle

\begin{tikzpicture}[remember picture,overlay]
    \node[anchor=north, yshift=-1cm] at (current page.north) {
        \fbox{\parbox{.99\textwidth}{\centering \small This work has been submitted to the IEEE for possible publication. Copyright may be transferred without notice, after which this version may no longer be accessible.}}
    };
\end{tikzpicture}

\begin{abstract}
\input{final_docs/abstract}

\end{abstract}


\input{final_docs/1_introduction}
\input{final_docs/2_related_work}
\input{final_docs/3_background}
\input{final_docs/4_ensemble_capacity}
\input{final_docs/5_problem_formulation}

\input{final_docs/6_solution_proposal}
\input{final_docs/7_eval_framework}

\input{final_docs/8_perf_eval}
\input{final_docs/9_lim_and_future_work}
\input{final_docs/10_conclusions}

\bibliographystyle{IEEEtran}
\bibliography{biblio}

\end{document}

%% file: acronyms.tex
\newacronym{bsm}{BSM}{Bell-state measurement}
\newacronym{cc}{CC}{classical communication}

\newacronym{dp}{DP}{dynamic programming}
\newacronym{dejmps}{DEJMPS}{Deutsch–Ekert–Josza–Macchiavello–Popescu–Sanpera}

\newacronym{ec}{EC}{ensemble capacity}
\newacronym{ep}{EP}{entangled pair}
\newacronym{edjp}{EDJP}{edge-disjoint paths}
\newacronym{edp}{EDP}{entanglement distribution problem}
\newacronym{eg}{EG}{entanglement generation}
\newacronym{egr}{EGR}{entanglement generation rate}
\newacronym{erp}{ERP}{entanglement routing problem}

\newacronym{gmcf}{GMCF}{generalized multi-commodity flow}

\newacronym{locc}{LOCC}{local operations and classical communication}
\newacronym{lp}{LP}{linear programming}

\newacronym{mcf}{MCF}{multi-commodity flow}

\newacronym{num}{NUM}{network utility maximization}

\newacronym{pmd}{PMD}{polarization mode dispersion}
\newacronym{povm}{POVM}{positive operator-valued measure}

\newacronym{qkd}{QKD}{quantum key distribution}
\newacronym{qn}{QN}{quantum network}
\newacronym{qsdc}{QSDC}{quantum secure direct communication}
\newacronym{qss}{QSS}{quantum secret sharing}

\newacronym{sdn}{SDN}{software-defined networking}
\newacronym{smo}{SMO}{service management and orchestration}
\newacronym{spdc}{SPDC}{spontaneous parametric down-conversion}

%% file: final_docs/abstract.tex
\rev{Large-scale quantum networks depend on the efficient distribution of entangled pairs between distant nodes. However, state-of-the-art solutions rely on restrictive operational assumptions, incur prohibitive computational complexity, and optimize metrics misaligned with application needs. This paper addresses the entanglement distribution problem introducing four pivotal advances. First, since the primary application of quantum communication is private information transmission, we derive the \textit{Ensemble Capacity}, a novel metric that quantifies the private classical information enabled by entanglement distribution. Second, we propose a generalized formulation of the \textit{Entanglement Distribution Problem} that allows arbitrary sequencing of entanglement swapping and purification, removing the structural restrictions of prior models. Third, to navigate the resulting combinatorial space, we introduce a dynamic programming-based hypergraph generation algorithm that preserves exact, continuous fidelities, avoiding quantization, while pruning sub-optimal sequences in advance. Finally, we integrate these solutions into \textit{CODE}, a two-tier orchestration framework enabling near-real-time network operation. Extensive evaluations show that our approach achieves higher private information capacity and significantly lower computational complexity than existing solutions, while meeting the sub-second latency requirements of dynamic quantum network operation.}

%% file: final_docs/1_introduction.tex
\section{Introduction}
\label{sec:introduction}
\noindent 

\Glspl{qn} are the next step in secure communications, as they leverage the principles of quantum physics to securely transmit data using entangled particles---typically polarized photons of light transmitted through optical fibers. \glspl{qn} enable new applications such as \gls{qkd}~\cite{travagnin2019quantum}, \gls{qss}~\cite{hillery1999quantum} and \gls{qsdc}~\cite{qi202115} \rev{which depend on efficiently generating and distributing \glspl{ep} between distant nodes to establish reliable, private end-to-end links.} However, distributing \glspl{ep} across long distances is far from trivial. Optical fiber losses grow exponentially with distance~\cite{pirandola2017fundamental}, and unlike classical signals, quantum states cannot simply be amplified or regenerated as a consequence of the no-cloning theorem~\cite{caleffi2018quantum}. 

Quantum repeaters~\cite{dur1999quantum, munro2015inside} \rev{overcome previous limitations through entanglement swapping~\cite{pan1998experimental}, which joins shorter entangled segments into longer ones. Yet imperfect repeater operations rapidly degrade} the \textit{fidelity}---the measure of an \gls{ep}'s proximity to its ideal reference state---\rev{so repeaters also perform entanglement purification~\cite{pan2001entanglement} to distill fewer high-fidelity pairs from several low-fidelity ones.} Swapping and purification operations are inherently interdependent: swapping extends distance but degrades fidelity, while purification increases fidelity at the cost of reducing the number of available \glspl{ep}. As a result, \glspl{qn} performance depends heavily on link management strategies that determine which repeaters (and in what order) execute these operations~\cite{victora2023entanglement, gu2023esdi, gu2024fendi}. In this work, we address the \textit{\gls{edp}}: the generation, transmission, and management of \acrlongpl{ep} across a QN. 

\begin{figure}[t!]
    \centering
    \setbox0=\hbox{\includegraphics[width=0.75\columnwidth]{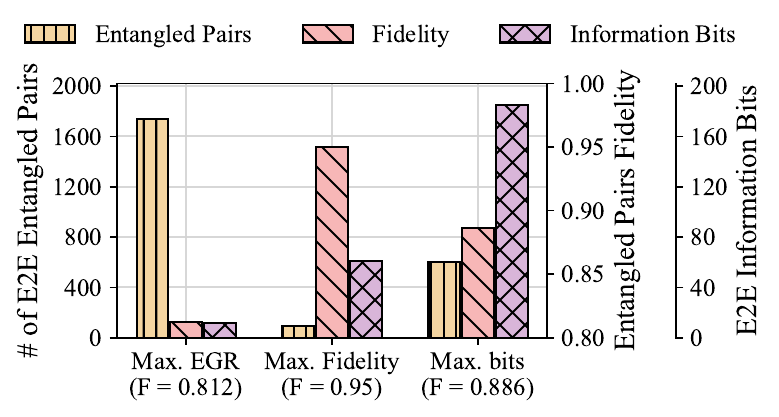}}
    \begin{minipage}[b]{0.27\columnwidth}
        \centering
        \begin{minipage}[b][\ht0][c]{\linewidth}
            \centering
            \includegraphics[width=\linewidth]{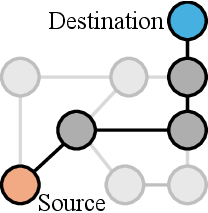}
        \end{minipage}
        \vspace{2pt}
        \centerline{\footnotesize (a)}
    \end{minipage}\hfill
    \begin{minipage}[b]{0.72\columnwidth}
        \centering
        \usebox0
        \vspace{2pt}
        \centerline{\footnotesize (b)}
    \end{minipage}
    \caption{Performance analysis of the proposed quantum network. (a) Schematic representation. (b) Comparison of different entanglement distribution strategies.}
    \vspace{-4mm}
    \label{fig:intro:combined}
\end{figure}

\rev{Existing solutions to the EDP either maximize the raw \gls{egr} (i.e., the rate at which end-to-end \glspl{ep} are shared) subject to hard fidelity constraints~\cite{fan2025distribution, caleffi2017optimal, pant2019routing, grant2023stochastic, gu2023esdi}, or maximize the end-to-end fidelity~\cite{gu2025cost, sun2024high} (see Sec.~\ref{sec:related}). However, neither objective necessarily maximizes the data actually transmitted, since both ignore the \textit{classical private information} capacity, which quantifies the amount of data that can be reliably transmitted. Fig.~\ref{fig:intro:combined} illustrates this limitation. Using our evaluation framework (see Sec.~\ref{sec:modeling_the_qn}) we have evaluated a 4-hop path within a QN assuming all links with the same distance (70 km), generation budget (2760 EPs per second), and initial fidelity. Using an exhaustive search, we computed the sequence of operations (i.e. swapping and purification) that maximize the EGR (\textit{Max. \gls{egr}}) and the end-to-end fidelity (\textit{Max. Fidelity}). Exhaustive search is only feasible for short length paths as the number of sequences grows super-exponentially with the path length (Sec.~\ref{problem_hardness}), which motivates the algorithms proposed in this paper. \textit{Max. \gls{egr}} delivers many pairs, but their low fidelity ($F = 0.812$) severely limits the transmission of private information. Conversely, \textit{Max. Fidelity} achieves a much higher fidelity ($F = 0.95$) but reduces the number of \glspl{ep} by $17\times$. Yet it transmits nearly $5\times$ more private bits than \textit{Max. \gls{egr}}, showing that neither rate nor fidelity alone determines the achievable private information.} Therefore, we introduce as the primary objective of the EDP the \acrfull{ec} (see Sec.~\ref{sec:on_the_capacity_of_a_qn}) which quantifies the actual secure classical bit rate enabled by distributed entanglement. Directly optimizing the EC (\textit{Max. bits}), strikes the optimal balance between rate and fidelity, achieving a 3x improvement in secure information bits over previous approaches. To the best of our knowledge, this work is the first to consider this dimension in the EDP.

\rev{Optimizing the EC requires a general solution space that allows arbitrary sequences of swapping and purification operations. Forcing purification to occur strictly before swapping~\cite{li2021effective, zhao2022e2e, chen2024optimum} leaves significant performance potential untapped. While unconstrained sequencing has been previously attempted using hypergraphs~\cite{fan2025distribution}, these models relied on the strict discretization of continuous fidelity values to keep the problem tractable leading to inherent degraded state accuracy and sub-optimal operational decisions. To overcome these limitations, we introduce a \gls{dp}-based hypergraph generation mechanism (Sec.~\ref{subsec:hypergraph_via_dp}) with two key advantages. First, it preserves the exact, continuous fidelity of the distributed EPs avoiding quantization errors. Second, it guarantees a finite and tractable solution space by identifying and pruning low-performing operational sequences in advance. Our evaluation shows that this approach greatly reduces computational with almost negligible accuracy loss, outperforming end-to-end performance compared to prior art.}

\rev{However, even an efficient hypergraph generation algorithm remains computationally heavy for the real-time-scale dynamics of a physical \gls{qn}. To bridge the gap, we introduce CODE (Sec.~\ref{sec:code}), a system-level orchestration framework with two nested loops. The \textit{Outer Loop} runs the computationally intensive steps on a coarse timescale (seconds to minutes), monitoring network conditions and demands. The \textit{Inner Loop} runs on a finer timescale (10~ms to 1~s), using the pre-computed structures to track rapid demand fluctuations and allocate resources in near-real-time. Our evaluation confirms that, unlike existing benchmarks, this hierarchical design meets the latency constraints of near-real-time \gls{qn} operation. To the best of our knowledge, CODE is the first system-level architecture that solves the EDP for practical deployment.}

We summarize the contributions of this work as follows:
\begin{enumerate}

    \item \textbf{Ensemble Capacity Metric:} We derive the \acrlong{ec} to quantify the private classical information that can be transmitted over a quantum channel and use it as the main optimization goal. 

    \item \textbf{Generalized Operational Sequencing:} We formulate a fully unconstrained mathematical model for entanglement distribution. \rev{Our solution unlocks previously unreachable regions of the optimization space.}

    \item \textbf{Continuous-Fidelity DP Hypergraphs:} We introduce a Dynamic Programming-based hypergraph generation algorithm that explores this space tractably by pruning low-performing branches, while preserving exact continuous fidelities instead of discretizing them.

    \item \textbf{System-Level Orchestration Framework:} We embed our solution in CODE, a dual-loop architecture that decouples heavy background optimization from near-real-time execution.

\end{enumerate}

%% file: final_docs/2_related_work.tex
\section{Related Work}
\label{sec:related}

Previous research on entanglement distribution has mainly focused on two different directions. On the one hand, many studies aim to maximize the end-to-end \gls{egr} without purification operations, which results in poor end-to-end fidelity, often using \gls{egr}-based routing metrics to select paths~\cite{caleffi2017optimal, pant2019routing, grant2023stochastic, gu2023esdi}. On the other hand, other works set a minimum end-to-end fidelity constraint to compute the best paths~\cite{chakraborty2019distributed, chakraborty2020entanglement, gu2024fendi} but neglect purification. \rev{Bridging both directions,~\cite{santos2023shortest} proposes a quasi-linear-complexity path-finding algorithm for \glspl{qn}, and~\cite{coutinho2023entanglement} extends this approach using \textit{fidelity curves}, which capture the rate--fidelity trade-off of each link, to jointly optimize end-to-end rate and fidelity through multi-objective path-finding. However, these works restrict purification to adjacent nodes and do not optimize its interplay with swapping.} Several works incorporated purification into path selection. The works \cite{li2021effective, zhao2022e2e, chen2024optimum} developed different strategies to ensure a minimum per-link fidelity through purification before swapping. \cite{li2022fidelity} formulates a joint routing–purification problem and designs algorithms for fidelity-guaranteed paths. In contrast, in this paper we incorporate both entanglement swapping and purification operations. More recently, the work more closely related to ours~\cite{fan2025distribution} considers both purification and swapping without constraints in the sequence of operations. This study aims to maximize the \gls{egr} under fidelity constraints. Nevertheless, as shown in Sec.~\ref{sec:introduction}, the maximization of the \gls{egr} does not necessarily correspond to the maximization of information transmission or \textit{classical private information}. Classical private information has also been studied analytically to bound secure key rates\rev{~\cite{pirandola2019end, harney2022endhybrid, harney2022end}}. \rev{While these works provide the theoretical capacity bounds for quantum repeaters and quantum networks for single and multi-path routing, they do not explicitly optimize entanglement distribution protocols.} The work in~\cite{victora2023entanglement} is the first to optimize entanglement routing using distillable entanglement, related to classical private information. However, it is limited to symmetric topologies, short paths, and shallow purification, providing a narrow perspective on the problem.

%% file: final_docs/3_background.tex
\section{Background}
\label{sec:background}

Quantum communication in quantum networks builds upon the following key principles:

\subsubsection{Quantum entanglement}

Entanglement~\cite{einstein1935can} is a phenomenon in quantum mechanics in which two or more particles share a global state that cannot be described by their local states. Measurements on one particle instantaneously correlate with outcomes on its entangled counterpart, a feature known as \emph{quantum non-locality}. Entangled pairs are the primary resource in \glspl{qn}, which generate, maintain, and distribute them across spatially separated nodes. The quality of an \acrlong{ep} is quantified by the \emph{fidelity} $F$, measuring the overlap between the actual state $\rho$ and a target Bell state $|\Phi^+\rangle$:
\begin{equation}
    F := \langle \Phi^+ | \rho | \Phi^+ \rangle,
\end{equation}
where $F = 1$ indicates a perfectly entangled pair, and values near $0$ indicate near orthogonality.
State-of-the-art techniques to generate entangled pairs include semiconductor quantum dots~\cite{huber2017highly}, nitrogen-vacancy centers in diamond~\cite{aharonovich2016solid}, atomic ensembles~\cite{duan2001long}, and nonlinear optical methods such as \gls{spdc}~\cite{smith2009photon}.

\subsubsection{Quantum teleportation and Entanglement swapping}
Quantum teleportation~\cite{bennett1993teleporting} is the fundamental mechanism for transferring a quantum state between distant parties without physical transmission, strictly respecting the no-cloning theorem~\cite{caleffi2018quantum}. 
Operating within the \gls{locc} framework, it requires a pre-shared \acrlong{ep}.
The sender performs a joint \gls{bsm} on the target qubit and their half of the entangled pair.
Upon receiving the two-bit classical measurement outcome from the sender, the receiver applies a conditional unitary operation to reconstruct the original state.

Entanglement swapping~\cite{pan1998experimental} leverages this exact mechanism to establish long-distance communication by effectively teleporting entanglement itself.
Given two independent pairs, $(A,B)$ and $(C,D)$, an intermediate node holding particles $B$ and $C$ performs a local BSM on them.
Once the intermediate node relays the classical measurement outcome to the remote end nodes for local corrections, particles $A$ and $D$ become entangled despite never physically interacting, seamlessly bridging the two initial segments.

\subsubsection{Entanglement purification}
Entanglement purification or distillation~\cite{bennett1996mixed}, is a probabilistic process designed to extract a smaller set of entangled pairs with higher fidelity from a large set of noisy or imperfectly entangled pairs~\cite{dur2007entanglement}.

%% file: final_docs/4_ensemble_capacity.tex
\section{Ensemble Capacity: Classical Private Information Analysis}
\label{sec:on_the_capacity_of_a_qn}

In this section, we introduce a novel metric to evaluate the performance of a \rev{quantum network}.
While metrics such as the end-to-end fidelity of shared pairs~\cite{gu2024fendi}, the end-to-end \acrlong{egr}~\cite{gu2023esdi}, the request fulfillment rate~\cite{zhang2025link}, and qubit utilization rate~\cite{pouryousef2023quantum} provide valuable insights, they do not directly quantify the network's data transmission capabilities. In contrast, we introduce the Ensemble Capacity to measure the volume of private classical information bits that two end parties can securely exchange. This metric inherently captures the trade-off between the quantity of shared entangled pairs (i.e. where the aggregate count is directly proportional to the \gls{egr}) and their fidelity. To address the entanglement distribution problem from a capacity-centric perspective (see Sec.~\ref{sec:problem_formulation}) we first define the capacity of a single Entangled Pair (see Sec.~\ref{single_pair_cap}) and then generalize this definition to a set of entangled pairs, formally defining the Ensemble Capacity (see Sec.~\ref{ensemble_cap}). Finally, we discuss the universality of this metric, making the Ensemble Capacity applicable across diverse quantum applications (Sec.~\ref{sec:metric_universality}). \rev{Main notation used in Sec.~\ref{sec:on_the_capacity_of_a_qn}--\ref{sol_proposal} is summarized in Table~\ref{tab:notation}.}

\begin{table}[t]
    \centering
    \caption{\rev{List of notations.}}
    \vspace{-2mm}
    \label{tab:notation}
    \footnotesize
    \setlength{\tabcolsep}{2pt}
    \renewcommand{\arraystretch}{1.05}
    \resizebox{.9\columnwidth}{!}{
        \begin{tabular}{@{}p{0.32\columnwidth} p{0.63\columnwidth}@{}}
        \hline
        \multicolumn{2}{@{}l}{\textit{Ensemble Capacity (Sec.~\ref{sec:on_the_capacity_of_a_qn})}} \\
        \hline
        $\rho$, $\{\rho_i\}_{i=1}^n$ & \gls{ep} and ensemble of $n$ EPs shared by two nodes \\
        $F$ & Fidelity of an \gls{ep} \\
        $\mathcal{C}(\rho)$ & Capacity of an \gls{ep} (bits) \\
        $H_2(\cdot)$ & Binary Shannon entropy \\
        $\mathcal{C}(\otimes_{i=1}^n \rho_i)$ & Ensemble Capacity, Eq.~\eqref{eq:capacity_of_a_q_link} \\
        \hline

        \multicolumn{2}{@{}l}{\textit{Entanglement Distribution Problem (Sec.~\ref{sec:problem_formulation})}} \\
        \hline
        $G=(V,E)$ & Graph: repeaters $V$, links $E$ \\
        $r_e$, $f_e$ & \gls{egr} (Hz), fidelity of link $e$ \\
        $T_w$ & Accumulation window \\
        $\mathcal{P}_{s,d}$ & Paths between $s$ and $d$ \\
        $\pi$, $\rho_\pi$& Entanglement Gen.\ Protocol and its EP ensemble\\
        $\mathcal{R}\pi, F\pi$, $T_p^{\pi}$ & Output EGR, fidelity, and processing time of $\pi$ \\
        $\mathcal{S}_P$, $\rho^{\mathcal{S}_P}$  & Entanglement Dist.\ Scheme and its product state \\
        $R^{e}_{\pi}$ & \gls{egr} of link $e$ used by $\pi$ \\
        \hline

        \multicolumn{2}{@{}l}{\textit{CODE solution (Sec.~\ref{sol_proposal})}} \\
        \hline
        $\mathcal{L}_P=(\mathcal{V},\mathcal{E})$ & Hypergraph of path $P$ \\
        $\mathcal{L}'_P$, $\mathcal{L}'_{\mathcal{P}_K}$ & Pruned hypergraph of $P$, $\mathcal{P}_K$ \\
        $\perp$, $\sqcup$ & Source, sink hyper-vertices \\
        $\epsilon=(L,l_{\text{out}},h,r_\epsilon)$ & Hyper-edge \\
        $h$ & \textit{swap}, \textit{purify}; aux.\ \textit{start}, \textit{end} \\
        $r_{\text{end}}$ & Rate of \glspl{ep} delivered to $d$ \\
        $L$, $l_{\text{out}}$, $r_\epsilon$ & Inputs, output, rate (Hz) of $\epsilon$ \\
        $f_0$, $f_{\text{swap}}$, $f_{\text{pur}}$ & Initial, swap, purify fidelity \\
        $\mathcal{F}$ & Discrete set of fidelity values \\
        \hline
        \end{tabular}
    }
    \vspace{-5mm}
\end{table}

\subsection{Single Entangled Pair Capacity}
\label{single_pair_cap}

We define the capacity $\mathcal{C}$ of a single \acrlong{ep} $\rho$ shared between a source $s$ and destination $d$ nodes as the amount of private classical bits $\mathcal{C}(\rho)$ that can be securely transmitted from $s$ to $d$ using $\rho$ assisted by adaptive \gls{locc} operations. 

Let $F$ denote the fidelity of an \gls{ep} $\rho$ modeled as a mixed two-qubit state (i.e., a Werner state modeling imperfect entanglement as a combination of a Bell state and white noise~\cite{askarani2021entanglement}). As established in~\cite{bowen2001teleportation}, teleportation over noisy entangled states, is equivalent to transmitting through a depolarizing quantum channel $\mathcal{E}_p$ of parameter $p=\frac{4(1-F)}{3}$, i.e. $\mathcal{C}(\rho) = \mathcal{C}(\mathcal{E}_p)$. Consequently as established in ~\cite{pirandola2017fundamental}, the capacity $\mathcal{C}(\rho)$ is bounded by  
\begin{equation}
k(p) - \frac{3p}{4}\log_2(3) \leq \mathcal{C}(\rho) \leq k(p),
\label{eq:bounds_on_capacity_p}
\end{equation}
where $k(p):=1-H_2(3p/4)$, and $H_2$ is the binary Shannon entropy. These bounds are derived assuming arbitrary teleportation protocols; this generality holds because QKD-style protocols can be reformulated in terms of entanglement-based teleportation, as shown in~\cite{bennett1992quantum}.

\subsection{Multiple Entangled Pairs Capacity}

\label{ensemble_cap}

Transmitting information via a single pair alone is often inefficient. Instead, entanglement is typically generated in parallel, producing multiple end-to-end pairs simultaneously. Therefore, we generalize the single-pair capacity to a network-level figure of merit: the \textit{Ensemble Capacity}. We define an ensemble as a set of \acrlongpl{ep} $\{\rho_i\}_{i=1}^n$ shared between a source $s$ and destination $d$. The \acrfull{ec}, denoted by $\mathcal{C}(\otimes_{i=1}^n\rho_i)$, is the number of private classical bits that can be securely transmitted between $s$ and $d$ by using the ensemble $\{\rho_i\}_{i=1}^n$ and adaptive LOCC operations.

While this definition captures the theoretical maximum data-carrying potential, computing $\mathcal{C}(\otimes\rho_i)$ is generally intractable due to the complexity of evaluating collective operations over entangled resources\rev{~\cite{qiqcop_op_d2813fe3fcdf09ad}}. Consequently, following the fundamental bounds in~\cite{pirandola2017fundamental}, we adopt a conservative approximation by considering a \textit{lower bound} as the sum of individual pair capacities, $\sum_{i=1}^n \mathcal{C}(\rho_i)$. 
\rev{\cite{leditzky2017quantum} suggests the gap to this bound to be small in the low-noise regime considered (see Sec.~\ref{sec:experimental_evaluation}).}

By substituting $p$ back into the lower bound of Eq.~\eqref{eq:bounds_on_capacity_p}, we obtain the computable lower bound for the Ensemble Capacity:
\begin{equation}
    \mathcal{C}(\otimes_{i=1}^n\rho_i) = \sum_{i=1}^n k\bigg(\frac{4(1-F_i)}{3}\bigg) - (1-F_i)\log_2(3).
    \label{eq:capacity_of_a_q_link}
\end{equation}
\subsection{Ensemble Capacity as a Universal Metric}
\label{sec:metric_universality}

\rev{The goal of quantum networks---regardless of the specific application---is to maximize the rate at which distributed quantum resources can be converted into desired target resources under the allowed operations. The relevant target, however, varies across use cases: for example, QKD maximizes secret classical bits, whereas distributed quantum computing and entanglement distribution maximize shared entangled pairs. This difference exposes a key limitation of the prevailing optimization methodology, which typically maximizes the entanglement generation rate subject to a hard fidelity constraint~\cite{gu2024fendi,zhao2022e2e, li2022fidelity}. Choosing a fixed threshold hard-codes the target application into the network design, yielding solutions that are not comparable across use cases \cite{kozlowski2020designing, jacinto2026network, perez2024simulation}, collapsing the trade-off between resource quantity and quality into an arbitrary operating point. In contrast, Ensemble Capacity provides a use-case-agnostic objective. It is based on a generic capacity that characterizes the asymptotic rate at which distributed quantum resources can be converted, under the allowed adaptive LOCC operations, into a desired target resource. Thus, depending on the target and operational task, it encompasses quantities such as distillable entanglement, two-way quantum capacity, and secret-key capacity, allowing the network to be optimized for maximal provisioning capability without fixing a particular application.}

%% file: final_docs/5_problem_formulation.tex
\section{Entanglement Distribution Problem formulation}
\label{sec:problem_formulation}

In this section, we present a mathematical framework for the Entanglement Distribution Problem that maximizes the \acrlong{ec} between a source $s$ and a destination $d$ by optimizing the underlying sequences of purification and swapping operations. Our formulation introduces three fundamental advancements over existing literature. First, it generalizes the selection of a single sequence of operations along a path $P$. Instead, our formulation allows multiple sequences to execute concurrently along the same path, generating an ensemble of distinguishable end-to-end EPs~\cite{victora2023entanglement, chen2024optimum}. Second, it allows purification operations at any hop of the distribution sequence. This flexibility significantly expands the solution space, unlocking high-fidelity strategies that are inaccessible to simpler models~\cite{li2021effective, zhao2022e2e}. Third, our formulation enables dynamic resource pooling across multiple, potentially overlapping paths.
By jointly optimizing over non-disjoint paths, the model recognizes when \glspl{ep} on shared links are underutilized due to bottlenecks elsewhere in the network. Consequently, our approach reallocates these residual resources to co-existing operational sequences, ensuring that no entanglement is wasted when chaining operations across the infrastructure~\cite{pant2019routing, chakraborty2019distributed}.

\subsection{Network Model and Notation}

A \acrlong{qn} is mathematically modeled as a graph $G = (V,E)$, where the vertex set $V$ represents the quantum repeaters and the edge set $E$ represents the physical links connecting adjacent nodes. Each edge $e \in E$ is characterized by the tuple $(u, v, r_{e}, f_{e})$, where $u$ and $v$ are two physically adjacent nodes, $r_e$ denotes the mean entanglement generation rate (in Hz) and $f_e$ represents the mean fidelity of the entangled pairs generated over the fiber-optic channel. We define the accumulation window $T_w$ as the time interval during which entangled pairs are accumulated in quantum memories. We assume $T_w$ is bounded by the quantum memory coherence time (see Sec.~\ref{subsec:qmemory_decoherence}). Let $\mathcal{P}_{s,d}$ be the set of all simple paths between source $s$ and destination $d$, where each path $P \in \mathcal{P}_{s,d}$ is an ordered sequence of adjacent physical links.

\subsection{Sequences of operations}

To establish an end-to-end entangled link between a source $s$ and a destination $d$ over a path $P$, $P$ must be transformed into a single virtual link through a structured sequence of entanglement swapping and purification operations. Establishing end-to-end connectivity for a path of length $k$ requires exactly $k-1$ swaps and provides $2(k-1)+1$ purification opportunities, as purification may be applied to the input links prior to each swap and to the terminal link. Valid sequences are defined by the specific allocation of purification rounds to these opportunities limited by the available resource pool.

\subsubsection{Entanglement Generation Protocols}

We define an entanglement generation protocol (EGP) $\pi$ as an ordered sequence of purification and entanglement swapping operations over the path $P$. The sequence of operations of a protocol $\pi$ transforms the path $P$ into a single virtual link 
generating an ensemble of indistinguishable EPs $\rho_\pi$ resulting in an end-to-end generation rate $R_\pi$ (in Hz) and fidelity $F_\pi$. A protocol $\pi$ operates on the batch of entangled pairs accumulated during a window $T_w$ processing them during a time $T^{\pi}_p$.

\subsubsection{Entanglement Distribution Schemes}

An entanglement distribution scheme $\mathcal{S}_P$ is defined by a set of entanglement generation protocols $\Pi = \{\pi_1, \dots, \pi_m\}$ and a corresponding resource allocation vector, enabling the concurrent execution of multiple protocols across path $P$. Formally, $\mathcal{S}_P = \{(\pi_j, \boldsymbol{R}_j)\}$, where $\boldsymbol{R}_j = (R^1_{\pi_j}, \dots, R^k_{\pi_j})$ specifies the absolute resource consumption for protocol $\pi_j$. Each $R^i_{\pi_j}$ denotes the raw EGR drawn from physical link $e_i \in P$ to support protocol $\pi_j$. A distribution scheme $\mathcal{S}_P$ effectively creates $m$ distinct ensembles of entangled pairs (i.e. $m$ distinct ``flows") yielding a product state $\rho^{\mathcal{S}_P}=\otimes^{m}_{i=1}\rho_{\pi_i}$ over a path $P$.
An empty scheme $\mathcal{S}_P = \emptyset$ indicates an idle path.

\subsection{Optimization Framework}

An optimal solution for the Entanglement Distribution Problem is rarely confined to a single sequence of operations over a single path. Since entanglement purification is a non-linear operation where investing additional resources yields diminishing returns, the optimal strategy often involves a composite solution. Consequently, the \gls{edp} must be optimized across multiple paths (utilizing disjoint or partially overlapping paths to aggregate capacity) and diverse EGPs executing heterogeneous sequences of operations). Thus, the challenge extends beyond simple shortest-path routing; it requires determining the optimal configuration of schemes $\mathcal{S}_P$ across the set of all simple paths $\mathcal{P}_{s,d}$ between $s$ and $d$ to maximize resource utilization while strictly adhering to link capacity constraints. 

We formulate the Entanglement Distribution Problem to maximize the aggregate capacity $\mathcal{C}$ between $s$ and $d$ as:

\vspace{-0.5cm}
\begin{subequations}
    \label{prob:entanglement_optimization}
    \begin{align}
        \max_{\{\mathcal{S}_P\}_{P \in \mathcal{P}_{s,d}}} \quad & \sum_{P \in \mathcal{P}_{s,d}} \mathcal{C}(\rho^{\mathcal{S}_P}) \label{prob:goal} \\
        \textbf{subject to:} \quad & \text{\underline{Resource Constraint:} } \notag \\
        &\displaystyle\sum_{P: e \in P}\sum_{\pi\in\mathcal{S}_P} R^{e}_\pi \leq r_e, \quad \forall e \in E. \quad \label{prob:const_cap} \\
        & \text{\underline{Non-negative Resource Consumption:} } \notag \\
        & \displaystyle R^{e}_\pi \geq 0 \quad  \forall e \in E, \quad \forall \pi \in \mathcal{S}_P.\label{prob:f_min}
    \end{align}
\end{subequations}

The objective function~\eqref{prob:goal} aggregates the capacities of all ensembles of entangled pairs generated across all active distribution schemes. The resource constraint~\eqref{prob:const_cap} ensures that the total physical consumption across all protocols using link $e$, denoted by $R^{e}_\pi$, does not exceed the available generation rate $r_e$. Finally, constraint~\eqref{prob:f_min} forces $R^{e}_\pi$ to non-negative values. 

Note that, while quantum entanglement generation is a discrete process (counting individual pairs), our framework models $R^{e}_{\pi_j}$ as continuous variables. This relaxation is justified by the operational context of the network. As the system operates over a large number of synchronized accumulation windows $T_w$, the discrete pair counts are aggregated into a long-term average throughput. Furthermore, the Ensemble Capacity $\mathcal{C}$ is an asymptotic metric defined over large ensembles of states. Therefore, treating $R$ as a continuous flow variable accurately represents the steady-state performance of the multiple distribution schemes.

\subsection{Problem Hardness and Intractability}
\label{problem_hardness}

The joint optimization of path selection and protocol configuration creates a nested combinatorial explosion that renders the problem intractable. First, the number of simple paths between a pair of nodes in a graph scales exponentially with the network size (specifically, as $\mathcal{O}(|V|!)$ in dense graphs~\cite{zhang2006introduction}). Second, for each candidate path $P$ of length $k$, the optimizer must configure a valid EGP $\pi$. On the one hand, the number of valid swapping operations grows following geometric permutations (specifically, as $\mathcal{O}(C_{k-1})$, where $C$ is the Catalan number~\cite{roman2015introduction} and $k$ is the number of links in the path), growing super-exponentially with path length. On the other hand, at each of the $2(k-1)+1$ purification opportunities, we must assign a specific number of purification rounds. Third, since the purification strategy on one link changes the consumption requirements for all other protocols sharing that link, these decisions cannot be made in isolation. They are coupled globally via the link capacity constraints (Eq.~\eqref{prob:const_cap}). Finally, as established in Sec.~\ref{sec:on_the_capacity_of_a_qn}, the capacity $\mathcal{C}$ is strictly convex with respect to fidelity. However, the process of purification dictates that fidelity $F$ and available rate $R$ of entangled pairs are inversely related via a non-linear efficiency curve. To increase $F$, one must non-linearly decrease $R$. Consequently, the interplay of massive combinatorial search spaces, globally coupled constraints, and non-linear physical dynamics renders exact optimization computationally infeasible.

%% file: final_docs/6_solution_proposal.tex
\section{CODE: Solution Proposal}
\label{sol_proposal}

To address the Entanglement Distribution Problem defined in the previous section, we propose a comprehensive optimization framework called \textbf{C}apacity \textbf{O}ptimization for \textbf{D}istributed \textbf{E}ntanglement (CODE). To manage the inherent computational intractability of the problem, CODE employs a hierarchical solution strategy. First, it extracts a subset of candidate paths, $\mathcal{P}'_{s,d} \subseteq \mathcal{P}_{s,d}$. Second, it determines the optimal configuration scheme, $\mathcal{S}_{P_i}$, for every path $P_i \in \mathcal{P}'_{s,d}$. 

The initial computation of $\mathcal{P}'_{s,d}$ mitigates the factorial complexity of path enumeration by restricting the search space to the $N$-shortest paths. The rationale behind this design decision, alongside the specific algorithmic details, is discussed in Sec.~\ref{sec:code}. Furthermore, to navigate the super-exponential solution space associated with $\mathcal{S}_{P}$, we adopt a hypergraph-based formulation. This abstraction effectively maps the combinatorial sequencing of quantum operations into a structured resource flow problem, enabling the application of Linear Programming (LP) techniques. 

Constructing this hypergraph, however, presents significant structural challenges. In Sec.~\ref{subsec:standard_hypergraph_formulation}, we introduce a standard hypergraph formulation prevalent in related literature, followed by a critical analysis of its limitations (Sec.~\ref{sec:hg_limitations}). Most notably, standard approaches rely on strict fidelity discretization to maintain a finite graph size. This coarse approximation introduces systemic inaccuracies and artificially degrades the achievable end-to-end network performance. 

To overcome these fundamental bottlenecks, we propose a novel \textit{DP-based Hypergraph Generation Algorithm} (Sec.~\ref{subsec:hypergraph_via_dp}). By leveraging dynamic programming, this approach efficiently generates a pruned hypergraph that circumvents the deleterious effects of fidelity discretization. Building on this refined structure, Sec.~\ref{subsec:LP_formulation} formulates the corresponding LP, while Sec.~\ref{subsec:complexity_analysis} provides a rigorous complexity analysis. Finally, Sec.~\ref{sec:code} synthesizes these algorithmic components to present a complete overview of the CODE architecture, alongside practical system implementation details.

\subsection{Hypergraph Formulation}
\label{subsec:standard_hypergraph_formulation}

\begin{figure}[t!]
    \centering
    \includegraphics[width=0.85\linewidth, trim={1.1cm 1.cm 1.1cm .4cm}, clip]{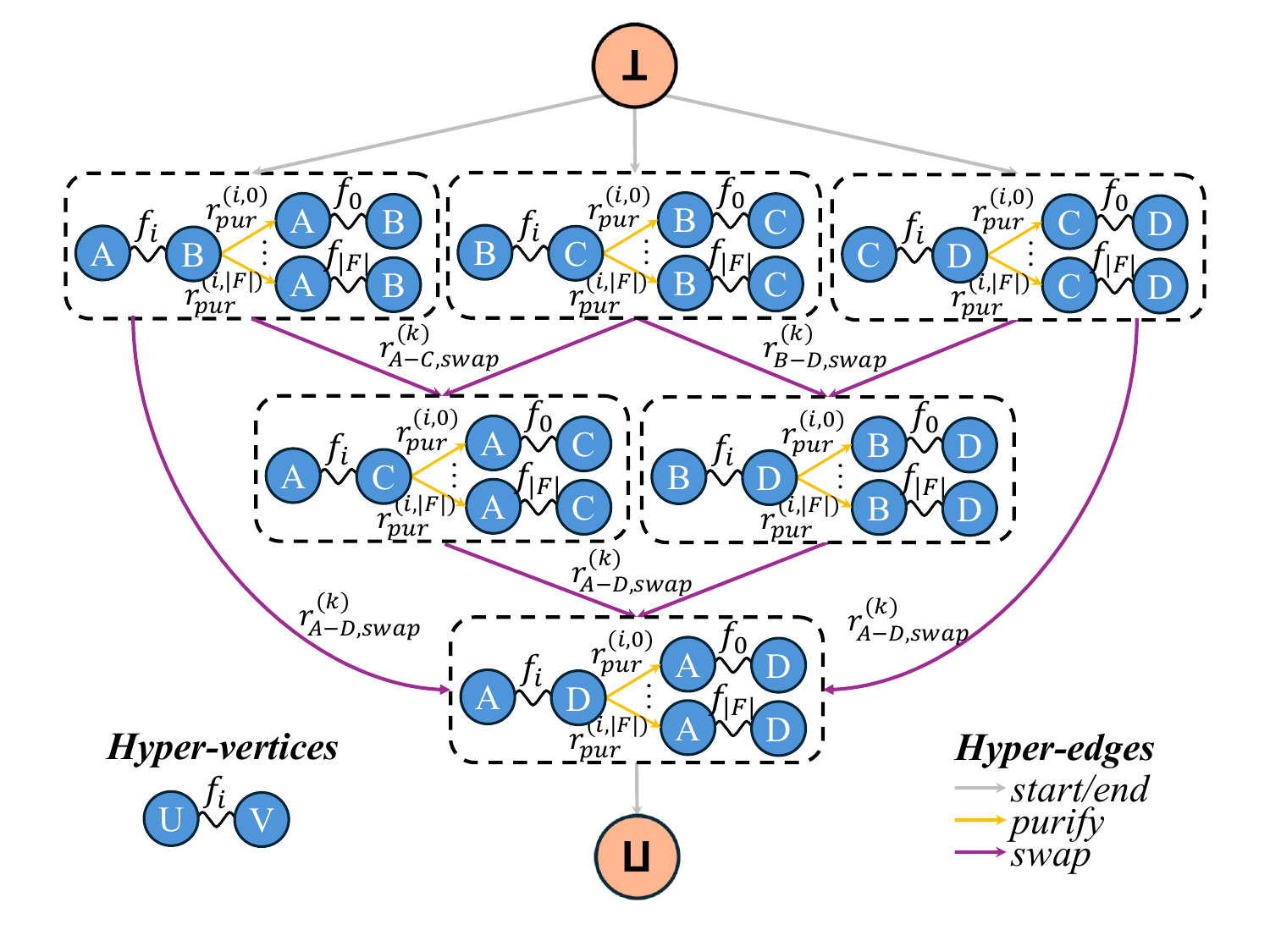}
    \vspace{-2mm}
    \caption{Example of the directed graph $\mathcal{L}_P$ for a path of 4 nodes and 3 links (A-B-C-D).}
    \label{fig:directed_graph_generation}
    \vspace{-3mm}
\end{figure}

To model the solution space, we build upon the state-of-the-art literature~\cite{fan2025distribution, gu2024fendi} to construct a directed hypergraph $\mathcal{L}_P=(\mathcal{V}, \mathcal{E})$ that encodes all feasible states of entanglement and the operational transitions between them. This abstraction maps the combinatorial complexity of the purification and swapping sequences into a navigable flow graph.

\subsubsection{Hyper-vertices (Entangled Links)}

The set of hyper-vertices $\mathcal{V}$ encodes the possible states of entangled links between two specific nodes within a path $P$. Each hyper-vertex is a tuple $(u,v,f)$, representing an ensemble of indistinguishable EPs with fidelity $f$ between nodes $u$ and $v$ on path $P$. If $u$ and $v$ are directly connected by an edge $e_{u,v} \in E$ in the physical topology, the entangled link is termed a \textit{physical link}. Otherwise, we term it as a \textit{virtual link}.
Additionally, we define two auxiliary hyper-vertices: the source vertex ($\perp$) which represents the initial pool of raw resources over the path and the sink vertex ($\sqcup$) which represents the final successful distribution of end-to-end EPs.

\subsubsection{Hyper-edges (Sequences of operations)}

The set of hyper-edges $\mathcal{E}$ models the transformation of a path $P$ through a sequence of swapping and purification operations. We define a hyper-edge $\epsilon \in \mathcal{E}$ as a transition $(L, l_{\text{out}}, h, r_\epsilon)$, where $L \subseteq \mathcal{V}$ is the set of input hyper-vertices, $l_{\text{out}}$ is the output hyper-vertex, and $h \in \{\textit{swap, purify, start, end}\}$ is the operation type and $r_{\epsilon}$ is the continuous consumption rate (Hz).

We distinguish between \rev{\textit{physical} operations} (swapping and purification, $|L|=2$, denoted as $l_1, l_2$), which \rev{are executed by the quantum nodes and} transform the state of existing entangled links, and \textit{auxiliary} operations (initialization and termination, $|L|=1$, denoted as $l_1$). \rev{Auxiliary operations are not executed by any hardware. We introduce them to formalize the hypergraph generation algorithm (Sec.~\ref{subsec:hypergraph_via_dp}) and the LP (Sec.~\ref{subsec:LP_formulation}), since they define where entanglement enters and leaves the flow graph.} Next, we formalize the former:
\begin{enumerate}
    \item \textbf{Swapping ($\textit{swap}$):} Combines two links sharing an intermediate node into one link between non-adjacent nodes. 
    \begin{equation}
        \begin{aligned}
            \textit{swap}\colon \mathcal{V} \times \mathcal{V} &\rightarrow \mathcal{V}\\
            (u,v,f_1), (v,w,f_2) &\longmapsto (u,w,f_{\text{swap}}(f_1,f_2))
        \end{aligned}
        \label{eq:sol_proposal:swap_def}
    \end{equation}
    Flow conservation requires equal consumption rate $r_{\epsilon}$ from both inputs. Residual rates on non-bottleneck links remain available in the resource pool for concurrent operations.
    
    \item \textbf{Purification ($\textit{purify}$):} Consumes two links between the same nodes to yield a single higher-fidelity link.
    \begin{equation}
        \begin{aligned}
            \text{purify}\colon \mathcal{V} \times \mathcal{V} &\rightarrow \mathcal{V}\\
            (u,v,f_1), (u,v,f_2) &\longmapsto (u,v,f_{\text{pur}}(f_1,f_2))
        \end{aligned}
        \label{eq:sol_proposal:purify_def}
    \end{equation}
    Note that this formulation supports both symmetric (where input fidelities are identical, $f_1=f_2$) and asymmetric purification (where  $f_1 \neq f_2$).
    
\end{enumerate}

The explicit functional forms of $f_{\text{swap}}$ and $f_{\text{pur}}$ are determined by the specific hardware implementation (see Sec.~\ref{sec:modeling_the_qn}). Additionally, we formalize the two auxiliary operations \rev{that define the boundaries of the flow}:

\begin{enumerate}
    \setcounter{enumi}{2}\item \textbf{Initialization ($\textit{start}$):} Maps the source vertex $\perp$ to a set of physical link $(u,v,f_0)$, representing raw entanglement generation: $(\{\perp\}, (u,v,f_0), \textit{start}, r_{\epsilon})$.
    \item \textbf{Termination ($\textit{end}$):} Maps end-to-end links $(s,d,f)$ to the sink vertex $\sqcup$, \rev{representing the decision not to operate further with those end-to-end \glspl{ep}}: $(\{(s,d,f)\}, \sqcup, \textit{end}, r_{\epsilon})$.
\end{enumerate}

\subsubsection{Hypergraph Structure}
\label{subsec:hypergraph_structure}
The directed hypergraph $\mathcal{L}_P=(\mathcal{V}, \mathcal{E})$ represents the operational solution space, where a connected sequence of hyper-edges from source $\perp$ to sink $\sqcup$ defines an entanglement generation protocol ($\pi$). To strictly model the physical constraints of entanglement swapping while ensuring that any spare capacity is preserved for other uses, we impose the following restrictions on the hypergraph:

\textbf{Pair-wise Flow Conservation}: A swapping hyper-edge consumes an identical flow rate $r_\epsilon$ from both input vertices. This reflects the physical reality that the end-to-end rate is strictly bounded by the link with the lower EGR.

\textbf{Resource Pooling}: Link vertices are modeled as cumulative resource pools. Multiple hyper-edges (representing distinct protocols) can draw from the same vertex. Physical validity is maintained by ensuring $\sum r_\epsilon \leq r_e$ for each link. This mechanism prevents the loss of ``spare capacity'' during asymmetric swaps, keeping residual EPs available for additional purification or parallel operations.

The Ensemble Capacity $\mathcal{C}$ for a protocol is then derived from the terminal flow $r_{\text{end}}$ and its associated fidelity $f$ at the sink vertex $\sqcup$, following the definition provided in Sec.~\ref{sec:on_the_capacity_of_a_qn}.

Fig.~\ref{fig:directed_graph_generation} depicts $\mathcal{L}_P$ for a four-node path $P=\{A, B, C, D\}$. The source vertex $\perp$ initiates entanglement distribution using physical links via the $\textit{start}$ operation. Dashed blocks represent fidelity manifolds for specific node pairs; within these, purification hyper-edges transition resource flow $r^{(i,j)}_{\textit{pur}}$ toward higher-fidelity states. Swapping hyper-edges utilize $r^{k}_{u-v, \textit{swap}}$ resources to execute pair-wise combinations of adjacent links: flow from $(A,B)$ and $(B,C)$ generates virtual links in $(A,C)$, which can then be combined with $(C,D)$ to establish the end-to-end link $(A,D)$. The ensemble capacity is computed as the final flow from the $(A,D)$ block is directed to the sink vertex $\sqcup$ through termination hyper-edges with EGR $r_{\text{end}}$. This configuration models all operational sequences as traversable paths within a unified flow graph.

\subsubsection{Hypergraph limitations}\label{sec:hg_limitations}

Modeling entanglement rates as continuous variables yields a continuous fidelity manifold $f \in [0.5, 1]$, resulting in a hypergraph $\mathcal{L}_P$ with an infinite vertex set. To tackle this problem, related literature~\cite{fan2025distribution} only considers a discrete set of fidelity values $\mathcal{F} = \{f_0, \ldots, f_{|\mathcal{F}|}\}$ for every entangled link in $\mathcal{L}_P$. They approximate the resulting link fidelity values $f$ to the largest $f_i\in \mathcal{F}$ s.t. $f\geq f_i$.

This approximation poses severe limitations. First of all, discretizing continuous fidelity values into predefined lower bounds forces physically distinct entangled pairs to share an identical fidelity representation. This uncertainty diminishes the overall system performance; since the capacity of the end-to-end ensemble is highly sensitive to fidelity, any inaccuracy in its measurement leads to sub-optimal configurations of the swapping and purification operations.

Second, discretization artificially degrades ensemble capacity by collapsing continuous fidelities into lower-bound intervals, rendering models insensitive to qualitative differences between paths. This creates a structural bias toward quantity over quality, where algorithms favor the highest raw entanglement rates while discarding lower-rate paths with superior true fidelities. Such sub-optimality compounds over multiple hops, systematically eliminating high-capacity trajectories. Furthermore, any attempt to mitigate this loss by increasing the resolution of the discrete set $\mathcal{F}$ triggers a curse of dimensionality, reintroducing the very computational intractability the approximation sought to avoid.

\subsection{DP-based Hypergraph generation algorithm}
\label{subsec:hypergraph_via_dp}

To address the fundamental limitations of the standard hypergraph formulation, we propose a novel hypergraph generation mechanism based on \gls{dp}. In contrast to prior approaches that force continuous quantum states into discrete values, our formulation explicitly preserves the exact, continuous fidelity values of the entangled pairs. Specifically, our DP-based algorithm constructs a pruned hypergraph that encapsulates only the most promising sequences of operations in terms of performance. By systematically filtering out sub-optimal configurations early in the process, this strategy effectively circumvents the curse of dimensionality, thereby substantially accelerating both the initial graph generation and the subsequent execution of the LP algorithm.

Crucially, this architecture relies on a decoupled optimization paradigm. The DP phase is designed exclusively to identify the maximum theoretical potential of isolated protocols. The subsequent LP formulation then leverages this solution space to determine the actual execution rates, globally optimizing the distribution schemes based on concurrent network demands and physical resource availability.

To systematically identify a set of high-performance protocols, direct maximization of the ensemble capacity cannot be employed, as this metric fails to satisfy the \textit{optimal substructure property} required by the DP formulation\footnote{Consider an optimal end-to-end ensemble generated via a final purification step. The predecessor states (i.e., two lower-fidelity ensembles shared between the same nodes) are not necessarily optimal in terms of their individual capacities. Consequently, a globally optimal capacity cannot be reliably constructed from locally optimal capacity subproblems, violating the optimal substructure property.} \cite{bertsekas2019reinforcement}.

To circumvent this fundamental limitation, we reformulate the DP objective: rather than optimizing capacity directly, the algorithm maximizes the achievable \gls{egr}, denoted by $r$, for a discrete set of minimum fidelities $f \in \mathcal{F}$. This reformulation provides two pivotal advantages. First, it restores the requisite optimal substructure property, rendering a problem strictly solvable via DP. Second, it yields a diverse, Pareto-optimal frontier of high-performance solutions, as the algorithm naturally identifies the optimal Entanglement Generation Protocol for every minimum fidelity tier defined in $\mathcal{F}$. Specifically, we employ a ``fidelity bucketing'' strategy. For each predefined fidelity interval $[f_i, f_{i+1})$ (where $f_i$ and $f_{i+1}$ are consecutive values in $\mathcal{F}$), the algorithm identifies the optimal state tuple $(u,v,f_{\text{exact}})$ where $f_{\text{exact}} \in [f_i, f_{i+1})$ such as the associated EGR is the maximum within the interval $[f_i, f_{i+1})$.

Crucially, unlike standard discretization techniques that systematically round continuous fidelities down to predefined thresholds, our fidelity-bucketing strategy retains the true continuous value, $f_{\text{exact}}$, within each hyper-vertex. This ensures that all subsequent operations during the hypergraph construction leverage precise physical parameters rather than pessimistic lower-bound approximations. Consequently, the algorithm rigorously preserves the most promising operational sequences while effectively pruning suboptimal trajectories, ultimately maintaining a finite and computationally manageable graph without sacrificing the physical accuracy of the underlying quantum states.

\begin{algorithm}[t!]
\caption{\\Hypergraph Construction via Dynamic Programming}\label{alg:dp}
\begin{algorithmic}[1]
\Require Network path $P$, set of physical links $\mathcal{V}_L$, physical link Entanglement Generation Rates ($\{r_l\}_{l=1}^{|P|-1}$), discrete fidelity levels $\mathcal{F}$, $\mathcal{L}'_P \leftarrow \emptyset$
\Ensure Pruned Hypergraph $\mathcal{L}'_P$
    \item[] Notation: $l_{u,v}^i\coloneq(u,v,f^{(i)})$, where $f^{(i)}\in[f_i,f_{i+1})$
    \State $\mathcal{V}'\gets \mathcal{V}' \cup \{\perp,\sqcup\}$ 

    \For{$l\in \mathcal{V}_L$}
        \State $\mathcal{V}'\gets \mathcal{V}' \cup  l$
        \State $\mathcal{E}'\gets \mathcal{E}'\cup (\{\perp\}, l, \text{start}, r_l)$
    \EndFor
    \For{$\Delta=2, \ldots, |P|-1$}
        \For{$u=1, \ldots, |P|-\Delta$}
            \State $v \gets u+\Delta$
             \For{$l_{u,w}^i,l_{w,v}^j\in\mathcal{V}'$} \Comment{Perform Swap Operations}
        \State $f_{\text{new}} \gets f_{\text{swap}}(l_{u,w}^i, l_{w,v}^j)$
        \State $r_{\text{new}} \gets \min(r_{e_{l_{u,w}^i}},r_{e_{l_{w,v}^j}})$
        
        \State Find $k$ s.t. $f_{\text{new}}\in [f_k,f_{k+1}]$
        \If{$r_{\text{new}} >r_{e_{l_{w,v}^k}}$}
            \State $\mathcal{V}' \gets \mathcal{V}' \cup l_{u,v}^{\text{new}}$
            \State $\mathcal{E}' \gets \mathcal{E}' \cup (\{l_{u,w}^i, l_{w,v}^j\},l_{u,v}^{\text{new}}, \text{swap}, r_{\text{new}})$
        \EndIf
    \EndFor
             \For{$l_{u,v}^i,l_{u,v}^j\in\mathcal{V}'$} \Comment{Perform Purify Operations}
        \State $f_{\text{new}} \gets f_{\text{purify}}(l_{u,v}^i, l_{u,v}^j)$
        \State $r_{\text{new}} \gets \min(r_{e_{l_{u,v}^i}},r_{e_{l_{u,v}^j}})\cdot p_{\text{succ}}$
        
        \State Find $k$ s.t. $f_{\text{new}}\in [f_k,f_{k+1}]$
        \If{$r_{\text{new}} > r_{e_{l_{w,v}^k}}$}
            \State $\mathcal{V}' \gets \mathcal{V}' \cup l_{u,v}^{\text{new}}$
            \State $\mathcal{E}' \gets \mathcal{E}' \cup (\{l_{u,v}^i, l_{u,v}^j\},l_{u,v}^{\text{new}}, \text{purify}, r_{\text{new}})$
        \EndIf
    \EndFor
        \EndFor
        \For{$f_i \in \mathcal{F}$}
            \State $\mathcal{E}' \gets \mathcal{E}'\cup (\{l_{s,d}^i\}, \sqcup , \text{end}, r_{l_{s,d}^i})$
        \EndFor
    \EndFor
\end{algorithmic}
\end{algorithm}

Algorithm~\ref{alg:dp} details the pseudocode for the proposed DP-based hypergraph generation framework. The procedure starts with an initialization phase (lines 1–5), wherein the physical links ($\mathcal{V}_L$) are mapped to the source vertex $\perp$ through the \textit{start} operation. Following this, the algorithm systematically explores all combinations of sub-paths within path $P$ via a set of nested iterations (lines 6–8). For each evaluated pair of nodes, the framework first considers entanglement swapping. It iterates over all available pairs of physical or virtual links (line 9) to establish a new end-to-end connection across an intermediate node. Upon computing the effective fidelity and EGR for a candidate swap (lines 10–11), the algorithm executes the pruning step: if the candidate yields a strictly superior EGR for its specific fidelity interval, the newly generated virtual link is registered in the hyper-vertex set $\mathcal{V}'$ (line 14), and its topological operation is appended to the hyper-edge set $\mathcal{E}'$ (line 15). Next, an analogous dynamic programming recurrence is executed to assess purification operations (lines 18–26). This step evaluates the distillation of available ensembles between the same node pairs, ensuring that all high-performing protocols are meticulously encoded into the final, computationally tractable hypergraph structure.
At last, the established end-to-end pairs are mapped to the end hyper-vertex $\sqcup$ through the \textit{end} operation (lines 28-30).

By strictly restricting the graph topology to these optimal operational trajectories, our methodology achieves a profound reduction in computational complexity. Specifically, the number of hyper-edges is compressed from $\mathcal{O}(|V|^3|\mathcal{F}|^2)$---the scaling characteristic of standard, unpruned formulations found in prior literature---to an upper bound of $\mathcal{O}(|V|^2|\mathcal{F}|)$. Crucially, this structural pruning is achieved without information loss \rev{associated to quantized fidelity}; the framework retains the exact, non-discretized fidelity parameters strictly required for the accurate computation of the Ensemble Capacity.

\subsection{Linear Programming Formulation}
\label{subsec:LP_formulation}

Using the previous pruned hypergraph $\mathcal{L}'_P=(\mathcal{V}, \mathcal{E})$, we formulate the Entanglement Distribution Problem presented in Sec.~\ref{sec:problem_formulation} as a \gls{lp} optimization problem. For each hyper-edge $\epsilon=(S, l_\text{out}, h, r_\epsilon)\in\mathcal{E}$, $r_\epsilon$ is the decision variable associated to the number of \acrlongpl{ep} generated by performing the corresponding operation $h\in\{swap, purify, start, end\}$. Let $\rho_\epsilon$ denote the quantum state of the \glspl{ep} generated following $\epsilon$. Then, the \acrshort{lp} formulation of the Entanglement Distribution Problem is:

\begin{subequations}
    \label{lp:formulation}
    \begin{align}
        \max_{r_\epsilon, \epsilon \in \mathcal{E}} \quad & \sum_{\epsilon \in \mathcal{E}_{\text{end}}} \mathcal{C}(\rho_\epsilon) \cdot r_\epsilon \label{lp:goal} \\
        \textbf{subject to:}& \text{ \underline{Flow Conservation:}} \quad \forall l \in \mathcal{V} \setminus \{\perp, \sqcup\}, \label{lp:flow_const}\\
        & \!\!\!\!\!\!\!\sum_{\epsilon \in in(l) \setminus \mathcal{E}_{\text{pur}}} \!\!\!\!\! r_\epsilon + \frac{1}{2} \!\!\!\! \sum_{\epsilon \in in(l) \cap \mathcal{E}_{\text{pur}}}\!\!\!\!p_{\text{succ}}(\epsilon) \cdot r_\epsilon \geq \!\!\!\sum_{\epsilon \in out(l)} \!\!\!\! r_{\epsilon}. \notag  \\
        & \text{ \underline{Generation Limits:}} \notag \\
        & \!\!\!\!\sum_{\epsilon \in \mathcal{E}_{\text{start}}(e)} r_\epsilon \leq r_e, \quad \forall e \in E. \label{lp:cap} \\
        & \text{ \underline{Non-negative Resource Consumption:}} \notag \\
        & r_\epsilon \geq 0, \quad \forall \epsilon \in \mathcal{E}. \label{lp:pos}
    \end{align}
\end{subequations}

The objective function \eqref{lp:goal} maximizes aggregate end-to-end Ensemble Capacity by summing terminal assigned execution rates $r_\epsilon$ weighted by their resulting state capacities $\mathcal{C}(\rho_\epsilon)$. Flow conservation \eqref{lp:flow_const} maintains entanglement balance for each vertex $l \in \mathcal{V} \setminus \{\perp, \sqcup\}$ by requiring that total incoming entanglement rate---comprised of non-purification flow ($in(l) \setminus \mathcal{E}_{\text{pur}}$) and purification yields ($in(l) \cap \mathcal{E}_{\text{pur}}$) scaled by $\frac{1}{2} p_{\text{succ}}(\epsilon)$ where $p_{\text{succ}}$ is the probability of successful purification---satisfies the total outgoing consumption in $out(l)$. This efficiency factor accounts for the physical sacrifice of two lower-fidelity pairs to produce one distilled pair. Generation limits \eqref{lp:cap} enforce Resource Pooling by restricting the aggregate execution of initialization protocols in $\mathcal{E}_{\text{start}}(e)$ to the hardware generation rate $r_e$ of physical link $e$. Finally, \eqref{lp:pos} mandates non-negative execution rates $r_\epsilon$ to ensure physical feasibility. \rev{Note that, in contrast to \cite{fan2025distribution}, there is no use-case-specific fidelity constraint.}

\subsection{Complexity Analysis}
\label{subsec:complexity_analysis}

Related literature shows that LP problems can be solved in polynomial time in the number of variables using interior-point methods~\cite{vanderbei1998linear}. Based on the formulation in~\eqref{lp:formulation}, the number of variables in our approach corresponds to the edge count $|\mathcal{E}|$ of the graph $\mathcal{L}_P$. For an unpruned hypergraph $\mathcal{L}_P$, $|\mathcal{E}|$ is on the order of $O(|V|^3|\mathcal{F}|^2)$, where $|V|$ represents the number of vertices in the path and $|\mathcal{F}|$ denotes the number of discrete fidelity values (Fidelity Resolution). While applying the pruning strategy reduces the variable count to $O(|V|^2|\mathcal{F}|)$, it incurs a computational cost. The DP pruning algorithm is recurrent, operating on $|V|^2$ states represented by all possible pairs of link nodes. Each state evaluates two branch types: "swap" and "purify". The purify branch requires checking $|\mathcal{F}|^2$ states, while the swap branch requires checking $|V| \cdot |\mathcal{F}|^2$ states. Consequently, the total complexity of the DP pruning algorithm is $O(|V|^3|\mathcal{F}|^2)$.

Despite the pruning algorithm having a complexity of $O(|V|^3|\mathcal{F}|^2)$, this pre-processing step is computationally advantageous. The runtime of LP solvers, denoted as $\mathcal{C}_{LP}(N)$, generally scales super-linearly with the number of variables $N$ (often $O(N^{3.5})$ or higher for interior-point methods). Without pruning, the total cost is dominated by solving the LP with a large variable set: $\mathcal{C}_{LP}(|V|^3|\mathcal{F}|^2)$. In contrast, our approach incurs the additive pruning cost of $O(|V|^3|\mathcal{F}|^2)$ but reduces the optimization step to $\mathcal{C}_{LP}(|V|^2|\mathcal{F}|)$.  Since the solver's complexity $\mathcal{C}_{LP}(N)$ grows significantly faster than linear time, the reduction in the dimensionality of the LP outweighs the overhead of the pre-processing algorithm.

\subsection{CODE: \textbf{C}apacity \textbf{O}ptimization for \textbf{D}istributed \textbf{E}ntanglement} \label{sec:code}

Having introduced the main building blocks of our solution, we unify the algorithmic components previously described into a cohesive orchestration framework termed CODE: \textbf{C}apacity \textbf{O}ptimization for \textbf{D}istributed \textbf{E}ntanglement. 

To effectively manage the trade-off between the computational cost of graph construction and the necessity for low-latency responsiveness to user demands, CODE operates via a hierarchical control plane composed of two nested loops (see Alg.~\ref{alg:code_framework}): the \textit{Outer Loop} and the \textit{Inner Loop}.
This architectural separation ensures that computationally intensive tasks whose results can be stored and reused further,  as long as the network state does not vary too much, are decoupled from real-time time-sensitive decisions.  

\subsubsection{The Outer Loop}
The Outer Loop operates on a coarse time scale (ranging from seconds to minutes) and serves as the network's topological manager. Its primary objective is to monitor global network conditions---such as physical link availability, background noise levels, and long-term traffic patterns---and update the routing tables accordingly. In this phase, the control plane executes the computationally heavy pre-processing steps: 

\begin{enumerate}
    \item \textbf{Path Enumeration:} It deploys a modified Dijkstra algorithm to extract a candidate set of the $N$-shortest paths \rev{in number of hops}, $\mathcal{P}'_{s,d} \subseteq \mathcal{P}_{s,d}$, for all active source-destination pairs $(s,d)$. \rev{We use the hop count because each additional hop requires one more entanglement swapping operation. As each swap degrades the fidelity (Sec.~\ref{sec:modeling_swapping}), paths with fewer hops have been shown to yield higher end-to-end fidelity and EGRs~\cite{li2021effective, li2022fidelity}}.
    
    \item \textbf{Protocol Optimization:} It executes the DP-based hypergraph generation algorithm (Sec.~\ref{subsec:hypergraph_via_dp}, Algorithm~\ref{alg:dp}) across all candidate paths $P \in \mathcal{P}'_{s,d}$ to identify the Pareto-optimal Entanglement Generation Protocols and evaluate the corresponding Ensemble Capacities of each protocol.
    
    \item \textbf{Topological Pruning:} It filters the candidate set by retaining only the top-$K$ paths ($\mathcal{P}_K \subseteq \mathcal{P}'_{s,d}$). In particular, we select the $K$ paths with the highest Ensemble Capacity of their best performing Entanglement Generation Protocols, discarding the $N-K$ underperforming paths.
    
    \item \textbf{Hypergraph Synthesis:} It aggregates the DP-optimized operational sequences derived from the selected $\mathcal{P}_K$ paths to construct and cache the global, multi-path pruned hypergraph, $\mathcal{L}'_{\mathcal{P}_K}$.
\end{enumerate}

\subsubsection{The Inner Loop}
The Inner Loop operates on a fine time scale ($10$ms to $1$s) and handles near-instantaneous user requests and rapid fluctuations in demand. Unlike the outer loop, which addresses structural optimization, the inner loop focuses purely on flow maximization. When a transmission request arrives, the Inner Loop retrieves the pre-computed pruned hypergraph $\mathcal{L}'_{\mathcal{P}_K}$ and solves the LP to determine the optimal Entanglement Distribution Scheme. 

This hierarchical approach offers a distinct computational advantage. By relegating the $\mathcal{O}(|V|^3|\mathcal{F}|^2)$ complexity of the DP-based pruning to the background Outer Loop, the critical path of the request processing is limited to solving the LP over a significantly reduced variable set. This ensures that the network can optimize entanglement distribution in near-RT without being bottlenecked by the combinatorial explosion of the full protocol space. 

\begin{algorithm}[t!]
\caption{\\ CODE: \textbf{C}apacity \textbf{O}ptimization for \textbf{D}istributed \textbf{E}ntanglement}
\label{alg:code_framework}
\begin{algorithmic}[1]
\Require Network topology $G$, noise sources (Sec.~\ref{sec:modeling_the_qn}), traffic patterns, $N$ (paths), $K$ (top paths),  $T_{outer}$ (Update period)
\Ensure Optimal Entanglement Distribution Scheme

\While{Network is active} \Comment{Outer Loop}
    \State Monitor physical link availability
    \For{each active source-destination pair $(s, d)$}
        \State $\mathcal{P}_N \leftarrow \text{runModifiedDijkstra}(G, s, d, N)$
        \For{each path $p \in \mathcal{P}_N$}
            \State $\{C_i\}_{i=1}^{N}\gets$ EC estimation via DP (Sec.~\ref{subsec:hypergraph_via_dp})
        \EndFor
        \State $\mathcal{P}_K \leftarrow \text{selectTopKPaths}(\mathcal{P}_N, \{C_i\}_{i=1}^{N}, K)$
        \State $\mathcal{L}'_{\mathcal{P}_K} \leftarrow \text{generatePrunedHypergraph}(\mathcal{P}_K)$
    \EndFor
    \State Store $\mathcal{L}'_{\mathcal{P}_K}$ for real-time access
    
    \While{$t_{\text{current}} < T_{\text{outer}}$} \Comment{Inner Loop}
        \If{Transmission request arrives}
            \State Retrieve stored hypergraph $\mathcal{L}'_{\mathcal{P}_K}$
            \State $S_{\mathcal{P}_K} \leftarrow \text{runLP}(\mathcal{L}'_{\mathcal{P}_K})$
            \State Execute $S_{\mathcal{P}_K}$
        \EndIf
    \EndWhile
\EndWhile
\end{algorithmic}
\end{algorithm}

%% file: final_docs/7_eval_framework.tex
\section{Evaluation Framework}
\label{sec:modeling_the_qn}

This section details the building blocks of our evaluation framework. This environment provides a unified reference to characterize how entanglement is generated and distributed across \acrlongpl{qn}. 
Developed in Python via the NetSquid~\cite{coopmans2021netsquid} library, the platform allows for the high-fidelity representation of scalable quantum networks. Unlike simplified models, our approach explicitly accounts for entanglement generation, swapping, and purification under realistic noise and hardware constraints.

\newsavebox{\figAbox}
\newsavebox{\figBbox}
\newlength{\figrowht}
\newlength{\panelAw}
\newlength{\panelBw}
\begin{figure*}[t!]
    \centering
    \setlength{\panelAw}{0.57\linewidth}%
    \setlength{\panelBw}{0.42\linewidth}%
    \sbox{\figAbox}{\includegraphics[width=1.0\panelAw]{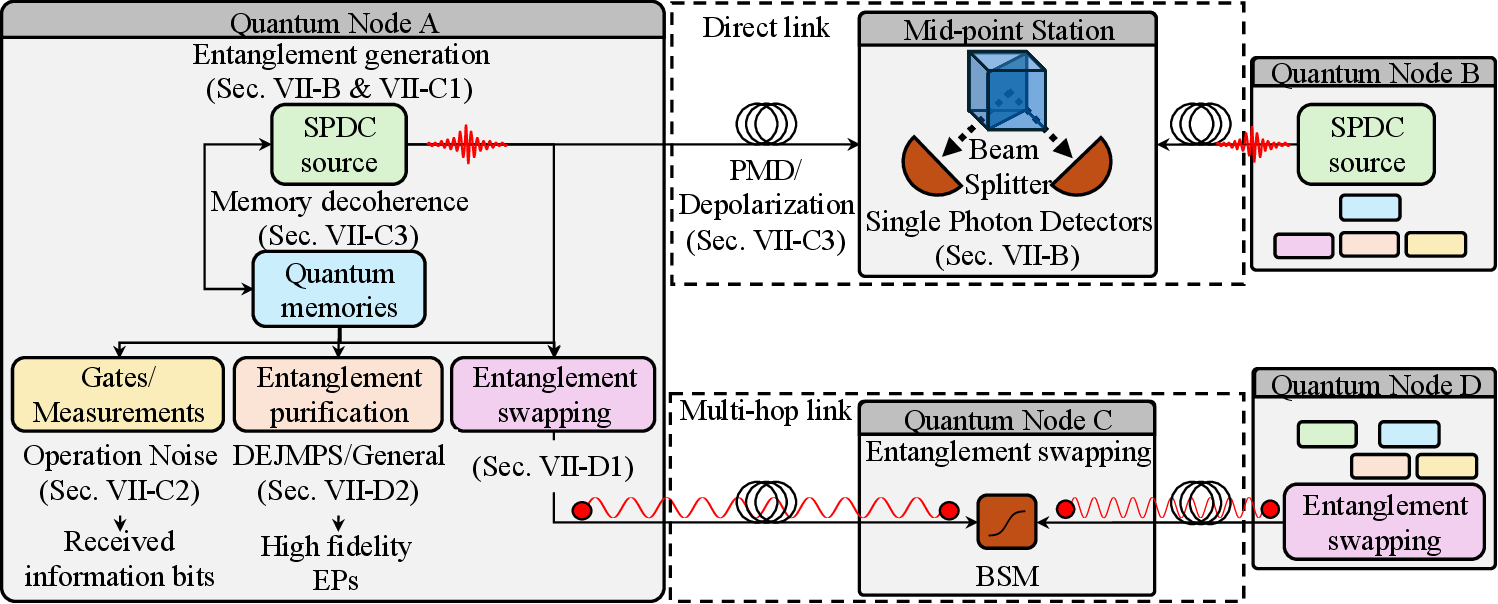}}%
    \sbox{\figBbox}{\includegraphics[width=1.0\panelBw]{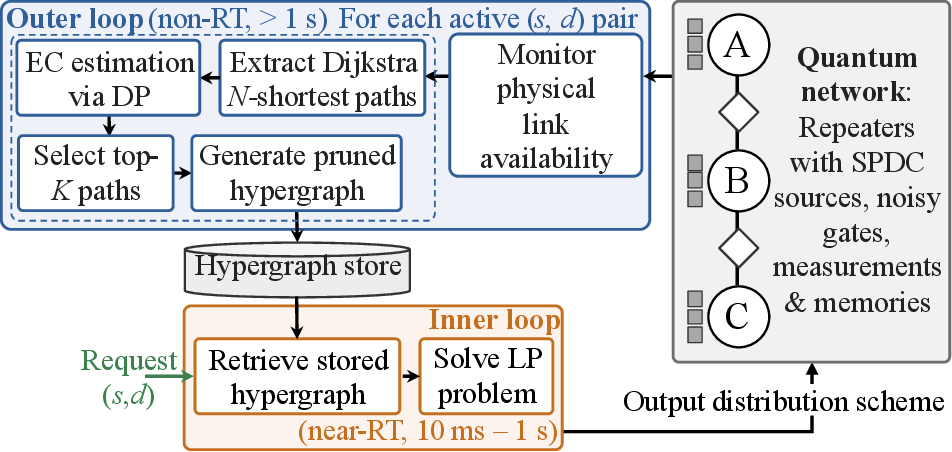}}%
    \setlength{\figrowht}{\ht\figAbox}%
    \ifdim\ht\figBbox>\figrowht \setlength{\figrowht}{\ht\figBbox}\fi
    \hfill
    \begin{subfigure}[b]{\panelAw}
        \centering
        \parbox[b][\figrowht][c]{\linewidth}{\centering\usebox{\figAbox}}
        \caption{Schematic representation of the \acrlong{qn} system model in Sec.\ref{sec:experimental_evaluation}.}
        \label{fig:system_model}
    \end{subfigure}
    \hfill
    \begin{subfigure}[b]{\panelBw}
        \centering
        \parbox[b][\figrowht][c]{\linewidth}{\centering\usebox{\figBbox}}
        \caption{\rev{Conceptual architecture of the CODE framework.}}
        \label{fig:code_architecture}
    \end{subfigure}
    \hfill\null
    \vspace{-3mm}
    \caption{Overview of the \acrlong{qn} system model and the CODE framework.}
    \label{fig:system_model_full}
\end{figure*}

\subsection{Quantum Network}
\label{net_model}

We model each node of the quantum network as a quantum repeater equipped with a \gls{spdc} source capable of generating entangled polarized photons encoding qubits. Quantum nodes can store qubits in quantum memories and perform computing operations via quantum gates and measurements on them, enabling Bell-State Measurements for quantum teleportation. Adjacent quantum nodes are connected via fiber-optic channels that enable the exchange of entangled photon pairs to establish entanglement through shared entangled pairs, enabling quantum teleportation of arbitrary states (Fig.~\ref{fig:system_model} \textit{Direct link}). Non-adjacent nodes can become entangled via sequential swapping operations across intermediate nodes, extending communication over long distances (Fig.~\ref{fig:system_model} \textit{Multi-hop link}). All these operations are subject to imperfections and noise. To mitigate any source of fidelity degradation, quantum nodes can also perform entanglement purification operations. Fig.~\ref{fig:system_model} shows the different building blocks of our \gls{qn} and the noise effects we have considered.

\subsubsection{Orchestration and Control Framework}
\label{subsubsec:control_framework}

A classical network overlaid on the QN interfaces with the quantum nodes using a centralized network controller. This controller enables the synchronization and coordination of quantum protocols. It is divided into two main control loops: ($i$) a non-real-time (non-RT) control loop that manages entangled resources and high-level policies (e.g. signaling link failures) with execution latencies exceeding 1s, and ($ii$) a near-real-time (near-RT) control loop which periodically reconfigures the quantum network and synchronizes the execution of different entanglement generation protocols with response times ranging from 10ms to 1s. \rev{Fig.~\ref{fig:code_architecture} shows diagram of the CODE framework (Sec.~\ref{sol_proposal}).} The controller ensures that the maximum processing time of all entanglement generation protocols $T^{\pi}_p$ is lower than the accumulation time window $T_w$. Furthermore, it ensures that $T^{\pi}_p +T_w$ does not exceed the quantum memories' coherence time (see Sec.~\ref{subsec:qmemory_decoherence}).

\subsection{Entanglement generation}
\label{subsec:modeling_entanglement_generation} 
To model entanglement generation, we adopt the mid-point link-level procedure presented in~\cite{zang2022simulation} and depicted in Fig.~\ref{fig:system_model_full} under the \textit{direct link} representation. Adjacent nodes store one photon from an SPDC-emitted pair and transmit the other to a mid-station. Successful entanglement is heralded by coincident detections after photons interfere at the mid-station's polarizing beam splitter.

At each quantum node, we model the effective local EGR at zero distance as $r_{\text{local}}$. This parameter accounts local hardware efficiencies including the probabilities of single-photon emission, memory-photon coupling and detector efficiency~\cite{hubel2007high, labay2024reducing, barends2014superconducting}. Consequently, the effective entanglement generation rate between adjacent nodes separated by a distance $L$ is determined by the fiber transmission loss as 
\begin{equation}
r_{\text{link}}(L) = r_{\text{local}} \cdot e^{-\alpha L / 10},
\end{equation}
\noindent where $\alpha$ represents the optical fiber attenuation coefficient. We used an attenuation coefficient $\alpha = 0.21 \text{ dB/Km}$~\cite{coopmans2021netsquid}.

\subsection{Entanglement Fidelity in Noisy Environments}
\label{sec:fidelity_noise}
The quality of a shared entangled pair can be quantified using its fidelity, which measures how close the actual state is to a target Bell state (see Sec. ~\ref{sec:background}). Several noise sources contribute to the degradation of entanglement fidelity.

\subsubsection{Initial Fidelity of Generated Entangled Pairs}
\label{subsec:noise_entanglement_transmission}
The initial fidelity of a two-photon state, measured immediately after generation with respect to an ideal Bell state and denoted as $F_0$, is typically observed to be slightly below unity due to imperfections in the generation process. Experimental results indicate that $F_0 \simeq 0.98$ is a conservative estimate for high-quality \gls{spdc} implementations~\cite{steinlechner2012high, magnitskiy2015spdc}.

\subsubsection{Imperfect Local Operations and Measurements}
\label{subsec:gate_decoherence}

Local operations such as quantum gates and measurements inevitably introduce inaccuracies and noise degrading entanglement fidelity and limiting communication performance. Also, these imperfections directly affect entanglement swapping and purification operations. We model these imperfections adopting the approach introduced in~\cite{dur1999quantum}:

\begin{itemize}
    \item \textbf{Quantum gate model}: We model a single-qubit gate noise using a parameter $p_1 \in [0,1]$, representing the lower bound on the probability of ideal operation. With probability $1 - p_1$, the noisy gate replaces the qubit with a maximally mixed state. Thus, the action of a noisy gate on a multi-qubit state $\rho$ is described as:
    
    \begin{equation}
    \vspace{-3mm}
        \rho \mapsto \underbrace{p_1 \, O^{\text{ideal}}_1 \rho}_{\text{Ideal operation}} + \underbrace{\frac{1 - p_1}{2} \, \operatorname{Tr}_1\{\rho\} \otimes \mathds{I}_1}_{\text{Depolarizing noise}},
    \end{equation}

where $O^{\text{ideal}}_1$ is the single-qubit ideal quantum gate and $\mathds{I}_1$ is the maximally mixed state on the affected qubit. Two-qubit gates are modeled analogously with reliability parameter $p_2$, and the depolarized component is replaced by a maximally mixed state on the involved qubits.

\item \textbf{Measurement model}: We model imperfect quantum state measurements using \glspl{povm}~\cite{helstrom1969quantum} with an accuracy parameter $\eta \in [0,1]$, representing the probability of correctly identifying the qubit state. For a single-qubit measurement in the computational basis $\{|0\rangle, |1\rangle\}$, the noisy measurement operators are defined as:
\vspace{-.2cm}
\begin{align}
    P_0^\eta &= \underbrace{\eta |0\rangle\langle0|}_{\text{Correct identification}} + \underbrace{(1 - \eta)|1\rangle\langle1|}_{\text{Misidentification}}, \\
    P_1^\eta &= \underbrace{\eta |1\rangle\langle1|}_{\text{Correct identification}} + \underbrace{(1 - \eta)|0\rangle\langle0|}_{\text{Misidentification}}.
\end{align}
\end{itemize}

In quantum teleportation, \glspl{bsm} project two qubits onto one of the four Bell states. We model \glspl{bsm} using an imperfect two-qubit CNOT gate followed by a Hadamard operation and two noisy single-qubit measurements as defined above. Experimental works in superconducting and spin-based quantum hardware~\cite{barends2014superconducting, dehollain2016optimization, rower2024suppressing, abughanem2024full} report reliability and accuracy parameters of $\eta = p_1 = p_2 = 0.995$.

\subsubsection{Decoherence in Quantum Memories}
\label{subsec:qmemory_decoherence}

Environmental noise degrades the fidelity of stored quantum states in quantum memories via two processes~\cite{wang2021single}: ($i$) thermal decoherence (i.e. amplitude damping), characterized by the relaxation time $T_1$, and ($ii$) dephasing, characterized by the coherence time $T_2$. Quantum states are assumed to remain coherent and unaffected by noise up to a fixed maximum storage time $t_{\text{cut}}$, beyond which they are considered fully decohered~\cite{victora2023entanglement, chen2024optimum, gu2024fendi}. This approximation is well-justified by the significant gap between the experimentally measured coherence durations \cite{abobeih2018one, bradley2019ten} ($T_1 = 3600$ s, $T_2 = 1.46$ s) and typical quantum protocol execution times on the order of $\sim10^{-2}$ seconds~\cite{dur1999quantum}.

\subsubsection{Depolarization in Optical Fiber}
\label{subsec:depolarization_from_fiber}

A polarization-encoded photon propagating through an imperfect optical fiber experiences a relative time delay denoted by $\tau_{\text{PMD}}$ between orthogonal polarization states. This phenomenon is referred as \gls{pmd}~\cite{hubel2007high} and is the primary source of fidelity degradation. When $\tau_{\text{PMD}}$ exceeds the photon's coherence time $\tau_{\text{coh}}$, polarization coherence is lost. 

We model $\tau_{\text{PMD}}$ as a normally distributed random variable ~\cite{hubel2007high, perez2024simulation}, $\tau_{\text{PMD}} \sim\mathcal{N}(DGD\cdot\sqrt{L},DGD\cdot\sqrt{L})$ where the $DGD$ is the differential-group delay in $\text{ps}/\sqrt{\text{Km}}$ and $L$ is its length of the fiber in Km. Experimental works~\cite{hubel2007high} estimate the $DGD$ of a single-mode fiber to be around $0.1 \text{ }\text{ps}/\sqrt{\text{Km}}$ and a coherence time $\tau_{\text{coh}}$ of $1.6 \text{ }\text{ps}$.

\subsection{Long-Distance Entanglement Distribution}
\subsubsection{Entanglement Swapping}
\label{sec:modeling_swapping}
Realistic swapping operations yield an entangled state of reduced fidelity due to imperfections and noise. After $N-1$ consecutive swapping operations across a $N$ hops, the resulting fidelity $F'$ of an EP is given by
\begin{equation}
F^{\prime} = \frac{1}{4} \left[ 1 + 3\left(p_1^2 p_2 \cdot \frac{4\eta^2 - 1}{3} \right)^{N-1} \prod_{i=1}^N \left( \frac{4F_i - 1}{3} \right) \right], 
\label{eq:swapping_fidelity}
\end{equation}

\noindent where $p_1$, $p_2$, and $\eta$ denote the reliabilities of the local operations and measurements and $F_i$ denotes the fidelity \rev{of the i-th physical link}
~\cite{askarani2021entanglement}. (We assume $p_1$, $p_2$ and $\eta$ equal for all network nodes in Eq.~\eqref{eq:swapping_fidelity}). This expression highlights that unless each operation and link is performed with near-perfect accuracy, the final fidelity $F^{\prime}$ decays exponentially with the number of swaps, asymptotically approaching $1/4$. This emphasizes the need of purification operations at intermediate nodes to enable long-distance entanglement distribution.

\subsubsection{Entanglement Purification}
\label{sec:modeling_purification}

\begin{figure}[t!]
    \centering
    \includegraphics[width=0.89\columnwidth, trim={8.5cm 4.8cm 9.cm 5.cm}, clip]{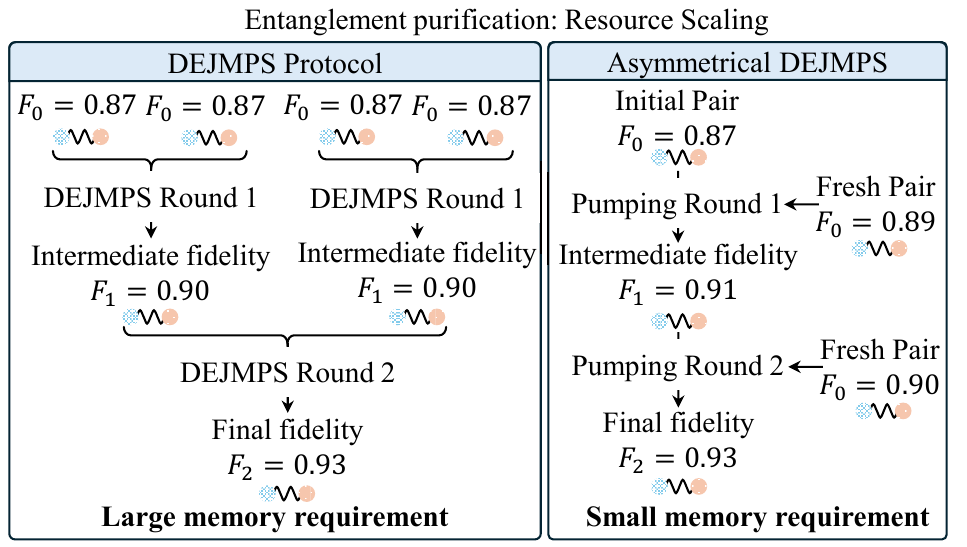}
    \caption{\gls{dejmps} and Asymmetrical \gls{dejmps} protocols.}
    \label{fig:resource_scaling}
\end{figure}

Entanglement purification is performed iteratively in discrete rounds, in which multiple low-fidelity pairs are combined to produce a smaller number of higher-fidelity pairs. Purification employs local quantum gates and measurements to decide whether the resulting pair is kept making the protocol inherently probabilistic. Fidelity gains diminish over successive rounds, asymptotically approaching a maximum attainable value \( F^{\text{max}}(p_1,p_2, \eta) < 1 \). In this work, we utilize the \gls{dejmps} protocol~\cite{deutsch1996quantum}, and extend it, via relaxing the identical-fidelity requirement, to an asymmetrical variant that allows for lower resource consumption.

\textbf{\gls{dejmps} Protocol}: The \gls{dejmps} purification protocol~\cite{deutsch1996quantum} operates on pairs of shared entangled particles with identical fidelities between two nodes (i.e. a transmitter and a receiver). Both apply to the pair local unitary rotations, followed by CNOT operations and a local measurement of one shared qubit in the computational $Z$ basis. Both nodes retain the remaining pair only if both parties obtain identical measurement outcomes. The protocol is fidelity-optimal for two copies of Bell diagonal states of rank up to three, achieving the maximum possible output fidelity and success probability~\cite{rozpkedek2018optimizing}. The protocol succeeds with probability (assuming $p_1$, $p_2$, and $\eta$ equal for both nodes)
\begin{equation}
    \label{eq:p_succ_dejmps}
    p_{\text{succ}}=\frac{1}{18}\left(9 + (4F_1-1)(4F_2-1)(1-2\eta)^2p_2^2\right),
\end{equation}
    
where $F_1$ and $F_2$ are the fidelities of the input pairs. The fidelity of the retained pair after successful purification is
\begin{equation}
\label{eq:dejmps_fidelity}
    \begin{split}
        F' &= \frac{9 +  p_2^2\left((1 - 8F_2)- 8F_1(6\eta^2 + 6\eta - 1)\right)}{72 \cdot p_{\text{succ}}} \\
        &+\frac{16 p_2^2F_1F_2(12\eta^2 - 12\eta + 5)}{72 \cdot p_{\text{succ}}}
    \end{split}
\end{equation}
\gls{dejmps}'s identical input fidelity constraint causes the expected number of consumed pairs to grow exponentially, scaling as $(2/p_{\text{succ}})^{k_{\text{max}}}$ for $k_{\text{max}}$ rounds. This exponential resource scaling (see Fig.~\ref{fig:resource_scaling}) poses a limitation in scenarios with constrained qubit memory.

\textbf{Asymmetrical \gls{dejmps}}: 
The asymmetrical variant is an adaptation of entanglement pumping~\cite{dur1999quantum} that relaxes the identical-fidelity requirement. It utilizes the same sequence of operations as the symmetrical case---and thus follows the same physical models for $p_{\text{succ}}$ and $F'$---but allows for $F_1 \neq F_2$.  Typically, a high-fidelity "protected" pair is iteratively purified by consuming a fresh, lower-fidelity "source" pair. While the single-step performance is upper-bounded by the symmetrical case, the asymmetrical approach allows for linear scaling of memory resources relative to the number of rounds.

%% file: final_docs/8_perf_eval.tex
\section{Performance Evaluation}
\begin{table}[t!]
\centering
\caption{Solution characterization across varying paths with capacity improvement relative to Rate-DP.}
\label{tab:performance_metrics}
\resizebox{\columnwidth}{!}{
\begin{tabular}{ll|ccc|ccc|c}
\toprule
 &  & \multicolumn{3}{c|}{\textbf{Performance Metrics}} & \multicolumn{3}{c|}{\textbf{Operational Metrics}} & Capacity  \\
\# & Strategy & EGR & Fidelity & Capacity & Swaps & Purif. & Pairs & Improv. \\
\midrule
\multirow{4}{*}{3} & Rate-DP & 2918.27 & 0.9266 & 1610.62 & 1.00 & 0.57 & 1.00 & --- \\
 & Rate-LP & \textbf{2920.20} & 0.9243 & 1584.53 & 1.20 & \textbf{1.70} & 1.20 & -1.62\% \\
 & EC-LP & \textbf{2920.20} & 0.9266 & 1614.89 & 1.78 & 1.24 & 1.78 & +0.27\% \\
 & \textbf{CODE} & \textbf{2920.20} & \textbf{0.9343} & \textbf{1738.62} & \textbf{1.80} & 1.41 & \textbf{1.80} & \textbf{+7.95\%} \\
\midrule
\multirow{4}{*}{5} & Rate-DP & 1902.40 & 0.8863 & 644.38 & 3.00 & 4.11 & 1.00 & --- \\
 & Rate-LP & 1971.85 & 0.8784 & 602.96 & 6.78 & \textbf{7.31} & 1.24 & -6.43\% \\
 & EC-LP & 2078.68 & 0.8800 & 666.59 & \textbf{9.46} & 6.16 & 3.19 & +3.45\% \\
 & \textbf{CODE} & \textbf{2163.37} & \textbf{0.8921} & \textbf{825.50} & 8.71 & 5.38 & \textbf{3.20} & \textbf{+28.11\%} \\
\midrule
\multirow{4}{*}{7} & Rate-DP & 704.13 & \textbf{0.8760} & 191.34 & 5.00 & 12.05 & 1.00 & --- \\
 & Rate-LP & 778.71 & 0.8701 & 185.70 & \textbf{19.41} & \textbf{21.78} & 1.01 & -2.95\% \\
 & EC-LP & 845.64 & 0.8678 & 194.29 & 18.99 & 19.05 & 3.36 & +1.55\% \\
 & \textbf{CODE} & \textbf{1117.58} & 0.8723 & \textbf{290.13} & 17.01 & 14.84 & \textbf{3.82} & \textbf{+51.63\%} \\
\midrule
\multirow{4}{*}{10} & Rate-DP & 276.11 & \textbf{0.8739} & 70.09 & 8.00 & 23.69 & 1.00 & --- \\
 & Rate-LP & 306.96 & 0.8700 & 72.60 & \textbf{39.67} & \textbf{51.99} & 1.00 & +3.58\% \\
 & EC-LP & 308.08 & 0.8709 & 73.21 & 38.11 & 48.74 & 3.50 & +4.44\% \\
 & \textbf{CODE} & \textbf{522.35} & 0.8721 & \textbf{127.15} & 28.99 & 35.36 & \textbf{3.90} & \textbf{+81.39\%} \\
\bottomrule
\end{tabular}}
\vspace{-10pt}
\end{table}

\label{sec:experimental_evaluation}
This section evaluates CODE against state-of-the-art solutions (Sec.~\ref{subsec:performance_benchmarking}, \ref{perf_bench}). The analysis covers Ensemble Capacity performance (Sec.~\ref{ec_opt_results}), operational optimization (Sec.~\ref{ec_opt_op}), and scalability (Sec.~\ref{ec_scale_path}). It assesses benchmark solution sensitivity to the minimum fidelity lower bound (Sec.~\ref{fidelity_low_b}), comparing optimal selections against CODE. Finally, it quantifies computational scalability (Sec.~\ref{subsec:scalability_analysis}), comparing CODE's execution times against benchmarks across varying fidelity discretization (Sec.~\ref{subsubsec:fidelity_discretization}) set sizes, path lengths (Sec.~\ref{subsubsec:effect_of_path_length}), and network sizes (Sec.~\ref{net_size_comp}). Our solution and benchmarks results use the evaluation framework described in Sec.~\ref{sec:modeling_the_qn}. We implemented CODE using Python, the evaluation framework using NetSquid, and we used Gurobi as the main solver for any LP formulation.

\subsection{Solution Benchmarks}
\label{subsec:performance_benchmarking}
State-of-the-art solutions to the Entanglement Distribution Problem do not optimize the Ensemble Capacity, rather they maximize the entanglement generation rate subject to minimum fidelity constraints. Thus, to fairly compare CODE against state-of-the-art solutions, we adapt them using a fixed fidelity lower-bound $f_{LB}$ and compute the resulting capacity using the end-to-end EGR. Selecting the optimal fidelity lower-bound $f_{LB}^*$ is non-trivial due to the complex, non-linear trade-offs between fidelity and generation rates (see Sec. \ref{perf_bench}). We compare CODE against the following strategies:

\begin{itemize}
    \item \textit{Rate-DP}: This solution is based on the dynamic programming formulation presented in~\cite{fan2025distribution}. It maximizes the EGR subject to a minimum fidelity constraint, restricted to entanglement pumping for purification.

    \item \textit{Rate-LP}: This solution is based on the LP formulation presented in~\cite{fan2025distribution}. It computes the hypergraph using the standard framework presented in Sec.~\ref{subsec:standard_hypergraph_formulation} and maximizes the EGR subject to a minimum fidelity constraint. 

    \item \textit{EC-LP}: This solution maximizes the Ensemble Capacity via the LP formulation presented in Sec.~\ref{subsec:LP_formulation} building a hypergraph using the standard framework presented in Sec.~\ref{subsec:standard_hypergraph_formulation}. This solution acts as an ablation study on the efficacy of DP-based hypergraph generation.
\end{itemize}

In order to get meaningful operational insights, we evaluate CODE and the benchmark solutions over realistic network topologies from TopoHub~\cite{jurkiewicz2023topohub}. This repository provides Gabriel graph models~\cite{cetinkaya2013fitness}, characterizing the proximity-based connectivity inherent in fiber-optic deployments, alongside empirical topologies from The Internet Topology Zoo~\cite{knight2011internet}.

\subsection{Performance Benchmarking}
\label{perf_bench}

We evaluate CODE and the benchmark solutions using a 1000-node Gabriel graph from the TopoHub repository~\cite{jurkiewicz2023topohub}. For each network link, distances are uniformly sampled from the range [20, 150] km. The local EGR at zero distance ($r_{local}$) is set to 12k entangled pairs per second (EPs/s)\cite{hubel2007high} for all nodes. For the \textit{Rate-DP} and \textit{Rate-LP} solutions, we fix the fidelity lower bound at $f_{LB} = 0.87$ (more details in Sec.~\ref{fidelity_low_b}). To ensure a fair comparison, all solutions were evaluated over a discrete set of fidelity values of size $|\mathcal{F}|=100$. We evaluate the solutions of different approaches across $100$ randomly selected node pairs for each path length (including both end nodes) of 3, 5, 7, and 10 nodes (corresponding to 2, 4, 6, and 9 hops, respectively).

Table~\ref{tab:performance_metrics} summarizes the averaged performance and operational metrics. Performance metrics include the mean end-to-end EGR, mean fidelity, and average Ensemble Capacity. Operational metrics encompass the average number of entanglement swaps and purification rounds required by the distribution scheme, alongside the mean number of distinguishable EPs produced. The final column reports the relative capacity improvement compared to the \textit{Rate-DP} benchmark. 

\subsubsection{Ensemble Capacity Optimization}
\label{ec_opt_results}

\begin{figure}[t!]
    \centering

    \begin{minipage}[t]{\columnwidth}
        \centering
        \includegraphics[width=0.85\columnwidth]{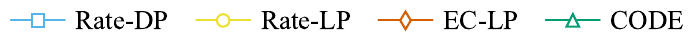}
        \begin{minipage}[t]{\columnwidth}

        \begin{subfigure}[b]{0.48\columnwidth}
            \centering
            \includegraphics[width=0.95\columnwidth]{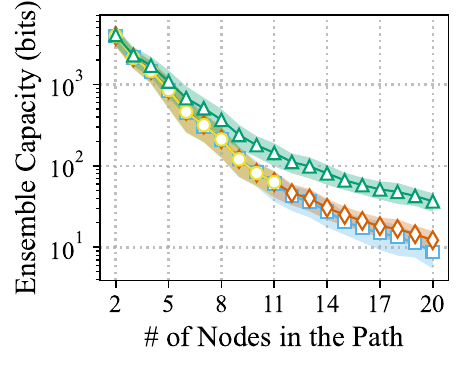}
            \caption{\small \acrlong{ec}}
            \label{fig:ec_vs_n_nodes}
        \end{subfigure}
        \hfill 
        \begin{subfigure}[b]{0.49\columnwidth}
            \centering
            \includegraphics[width=0.95\columnwidth]{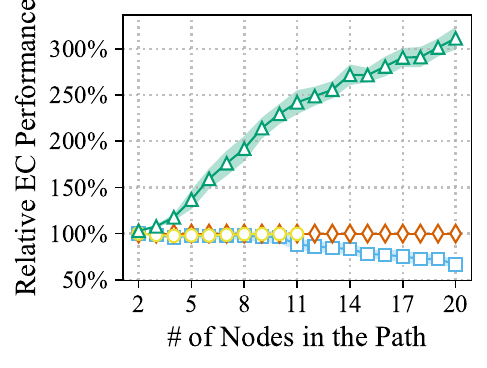}   
            \caption{\small Relative EC Gains}
            \label{fig:normalized_ec_vs_n_nodes}
        \end{subfigure}       
                    
        \end{minipage}
        \vspace{-0.25cm}
        \caption{Comparison of the \acrlong{ec} and Relative \acrlong{ec} Gains as a function of the path length.}
        \label{fig:performance_benchmarking_2}
    \end{minipage}
\end{figure}

The results presented in Table~\ref{tab:performance_metrics} show that CODE achieves better Ensemble Capacity optimization across all path lengths. While it provides a modest $7.95\%$ improvement in 3-hops paths, its improvement scales dramatically to a $81.39\%$ capacity increase in 10-hop paths. Also, CODE achieves the highest end-to-end EGR and top-2 fidelity across all path lengths. Since CODE utilizes the discretization set $\mathcal{F}$ as interval boundaries, it accesses a richer search space of entanglement distribution schemes, identifying high-potential solutions that standard methods would otherwise prematurely prune.

On the other hand, \textit{EC-LP} consistently delivers higher EGR than both \textit{Rate-DP} and \textit{Rate-LP} as it aggressively uses residual resources to maximize the Ensemble Capacity. Conversely, despite optimizing a superset space of \textit{Rate-DP}\cite{fan2025distribution}, \textit{Rate-LP} unexpectedly underperforms \textit{Rate-DP} in capacity for medium-to-short chains. In these scenarios, \textit{Rate-LP} suffers from a ``high-rate, low-fidelity'' trade-off; the solver selects high EGR solutions that are undermined by poor end-to-end fidelity.

\subsubsection{Operational Optimization}
\label{ec_opt_op}
Operationally, CODE generates a highly heterogeneous ensemble of end-to-end entangled pairs. By optimizing residual resources, it maximizes EPs yield while maintaining the number of swap and purification operations comparable to benchmark solutions. While \textit{EC-LP} also produces high-utility heterogeneous ensembles, it achieves lower total capacity than CODE. Conversely, as \textit{Rate-DP} is constrained by a single distribution protocol, it limits purification opportunities and ties swap counts to path length. \textit{Rate-LP} is operationally inefficient, performing additional swapping and purification operations that fail to increase the ensemble size significantly beyond \textit{Rate-DP} levels.

\subsubsection{Scaling with Path Length}
\label{ec_scale_path}

To evaluate solution scalability, we analyze performance as path length increases. We now use the real link distances of the 1000-node Gabriel graph. Link distances range from 25 to 400 km with a mean 100 km. As previously established, we randomly select node pairs spanned by defined path lengths. Figure~\ref{fig:ec_vs_n_nodes} presents the 95\% confidence interval for Ensemble Capacity computed across 20 distinct paths per length size as a function of node count. To isolate algorithmic efficiency from path-dependent variables (e.g., total length, entanglement generation rate, and intermediate link fidelity), Figure~\ref{fig:normalized_ec_vs_n_nodes} displays the Relative EC Performance. This secondary analysis normalizes each method against the \textit{EC-LP} benchmark on a per-path basis. \textit{EC-LP} serves as the 100\% baseline due to its consistently superior performance among state-of-the-art methods. For \textit{Rate-LP} solution we cannot use paths with more than length 10 as we exhaust our available memory to run the solution.

We observe that the absolute Ensemble Capacity of all four solutions decreases as the path length grows. This behavior aligns with physical expectations: longer paths require more entanglement swapping and purification operations inevitably leading to accumulated fidelity degradation and lower effective EGR across the end-to-end connection. 

CODE consistently outperforms the other methods across all path lengths, yielding the highest overall capacity and exhibiting the most resilience to network scaling. 
While it shows a modest improvement over the \textit{EC-LP} at a 2-node configuration, its relative advantage scales steadily, achieving over a $300\%$ performance gain at 20 nodes path length.

\subsection{Fidelity Lower-Bound Sensitivity}
\label{fidelity_low_b}
\begin{figure}[t!]
\centering
\begin{minipage}[t]{\columnwidth}
\centering
\includegraphics[width=.8\columnwidth, trim={-.5cm 0.4cm 0.4cm 0.3cm}, clip]{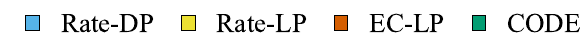}

    \begin{minipage}[t]{\columnwidth}

    \begin{subfigure}[b]{0.48\columnwidth}
        \centering
        \includegraphics[width=0.95\columnwidth]{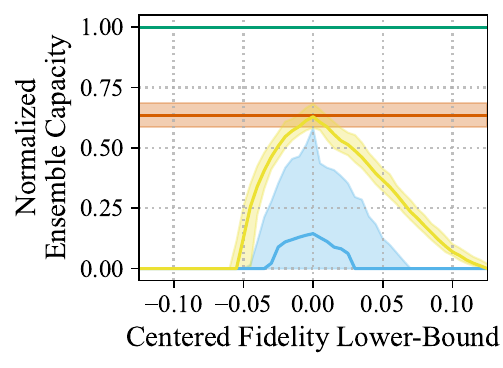}
        \caption{\small Normalized EC}
        \label{fig:normalized_capacity_vsnormalized_threshold}
    \end{subfigure}
    \hfill
    \begin{subfigure}[b]{0.49\columnwidth}
        \centering
        \includegraphics[width=0.87\columnwidth]{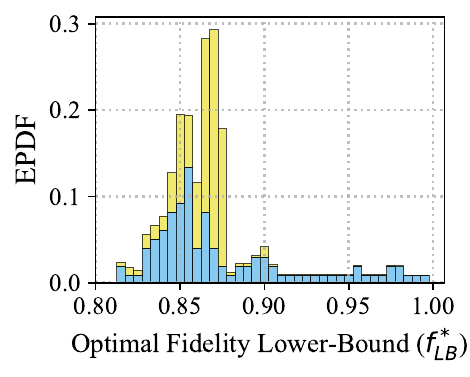}
        \caption{\small $f^*_{LB}$ Distribution}
        \label{fig:histogram_maxima}
    \end{subfigure}
                
    \end{minipage}
    \vspace{-5pt}
    \caption{Impact of the $f_{LB}$ on network performance.}
    \label{fig:ec_vs_threshold}
\end{minipage}
\vspace{-5pt}
\end{figure}

\begin{figure*}[t!]
    \centering

    \includegraphics[width=0.4\linewidth]{final_figures/plot_legend.pdf}
    
    \begin{minipage}[t]{0.32\linewidth}
        \centering
        \includegraphics[width=0.8\columnwidth, trim={0cm 0.2cm 0.cm 0.0cm}, clip]{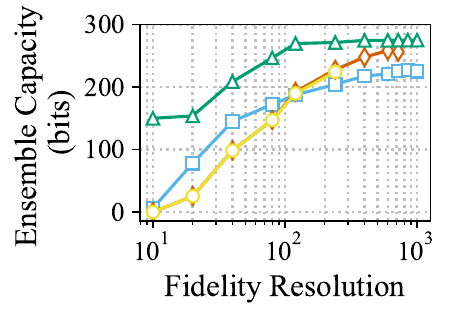}
        \vspace{.0005cm}
        \caption{Ensemble Capacity as a function of the Fidelity Resolution.}
    \label{fig:capacity_vs_granularity}
    \end{minipage}
    \hfill
    \refstepcounter{figure}
    \raisebox{-0.25cm}{
    \begin{minipage}[t]{0.65\linewidth}    
        \begin{subfigure}{0.49\columnwidth}
        
            \centering
            \includegraphics[width=0.82\columnwidth]{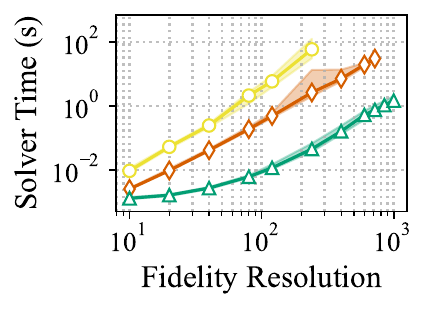}
            \caption{Solver Time}
            \label{fig:solver_time_vs_granularity}
        \end{subfigure}
    \hfill
        \begin{subfigure}{0.49\columnwidth}
            \centering
            \includegraphics[width=0.82\columnwidth]{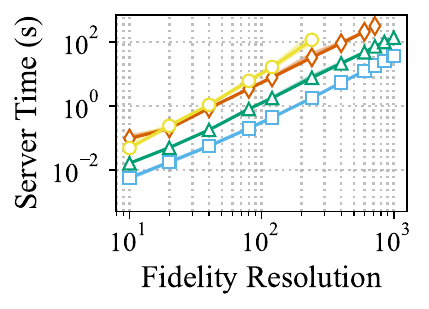}
            \caption{Server Time}
            \label{fig:server_time_vs_granularity}
        \end{subfigure}
        \vspace{2pt}
        \addtocounter{figure}{-1}
        \caption{Running time analysis as a function of the Fidelity Resolution.}
    \label{fig:time_vs_granularity}
    \end{minipage}}

\end{figure*}

We now evaluate the sensitivity of benchmark's results to fidelity lower bound constraint. Even though CODE does not depend on the fidelity lower bound constraint as it optimizes directly the Ensemble Capacity, it is helpful to study how sensitive state-of-the-art solutions are to this contraint as its value largely affect the quality of solutions. To study this dependence, we resort to the Gabriel graph used in Sec~\ref{ec_opt_results}, and run the CODE and the benchmark solutions for $f_{LB}$ values between 0.815 and 0.995 in 0.005 increments. We analyze the results for 100 randomly selected pairs of nodes connected between a path with lengths between 2 to 10 hops. Fig.~\ref{fig:normalized_capacity_vsnormalized_threshold} plots the normalized Ensemble Capacity metric as a function of the centered fidelity lower-bound, $f^c_{LB} = f_{LB} - f^*_{LB}$ where $f^*_{LB} = 0.87$ maximizes the Ensemble Capacity for \textit{Rate-DP} and \textit{Rate-LP}. Centering the fidelity decouples intrinsic protocol performance from the path-dependent optimal operating points of the benchmarks. Per-path normalization of the Ensemble Capacity values isolates topological variability, allowing for a fair cross-scenario comparison. Also, CODE's capacity defines the EC normalization baseline.

As we observe in Fig.~\ref{fig:normalized_capacity_vsnormalized_threshold}, CODE directly optimizes the Ensemble Capacity without a fixed fidelity constraint, its performance is strictly invariant to $f_{LB}$. \textit{EC-LP} solution is the closest benchmark to CODE, achieving approximately 65\% of CODE's capacity while also remaining invariant to $f_{LB}$. This performance gap highlights the limitations inherent in the standard hypergraph models utilized by \textit{EC-LP}. By avoiding the rigid fidelity discretization characteristic of prior art, CODE accesses a richer search space of Entanglement Distribution Schemes, exploiting high-potential solutions that standard discretization methods prematurely prune.

Conversely, the \textit{Rate-DP} and \textit{Rate-LP} approaches exhibit a distinct Gaussian-like sensitivity to $f_{LB}$. Their performance depends heavily on the precise optimization of $f_{LB}$, a non-trivial task dependant on multiple path-specific factors, including link count, link-level fidelities, and EGRs. Consequently, no single fidelity lower-bound is universally optimal across diverse network configurations. Fig.~\ref{fig:histogram_maxima} presents the distribution of optimal values ($f^*_{LB}$) for the \textit{Rate-DP} and \textit{Rate-LP} solutions across multiple Gabriel topologies and path lengths. Although the $f^*_{LB}$ values are distributed within the interval $(0.8, 1.0)$, benchmark performance exhibits strong sensitivity to the exact selection of $f_{LB}$. A marginal deviation of 0.02 from $f^*_{LB}$ yields up to a 20\% degradation in Ensemble Capacity.

\subsection{Computational Scalability Analysis}
\label{subsec:scalability_analysis}
Finally, we evaluate the computational scalability of CODE against the stat-of-the-art solutions. We analyze the execution time as a function of the fidelity resolution---defined as the discretization set cardinality, $|\mathcal{F}|$--- and the total number of nodes in the path and network. Quantifying these computational trade-offs delineates the operational limits of the benchmark formulations and validates CODE's applicability for large-scale quantum network deployments.

\subsubsection{Fidelity Discretization}
\label{subsubsec:fidelity_discretization}

To begin with, we evaluate the sensitivity of the execution time of the different solutions to the fidelity resolution. As explained in Sec.~\ref{sol_proposal}, $|\mathcal{F}|$ sets the size and precision of the state space during the hypergraph construction. A higher resolution allows for a more accurate tracking of entanglement states, which enables the discovery of higher-yield routing solutions. However, this comes at the cost of a larger state space, increasing the computational burden (see Sec.~\ref{subsec:complexity_analysis}). We evaluate the impact of Fidelity resolution using a 1000-node Gabriel graph from TopoHub repository~\cite{jurkiewicz2023topohub}. We now fix the path length to 6 nodes and study the effect of the fidelity resolution in the output EC.

Figure~\ref{fig:capacity_vs_granularity} depicts the Ensemble Capacity as a function of fidelity resolution. As expected, all solutions exhibit an increase in capacity as the resolution becomes higher. The state-of-the-art benchmarks start with near-zero capacity at low resolution and require highly refined state spaces to identify viable entanglement distribution schemes. In contrast, CODE demonstrates remarkable algorithmic efficiency at low resolutions. It achieves near-peak capacity at a resolution of just $10^2$, while the competing methods must scale toward $10^3$ to reach comparable, albeit still inferior, performance levels. 

Fig.~\ref{fig:solver_time_vs_granularity} presents the Solver Time which is time required for the LP solver to find the optimal solution while fig.~\ref{fig:server_time_vs_granularity} presents the Server Time which is time required to construct the state space and optimization model. Both plots display the 1st, 50th, and 99th percentiles. \textit{Rate-DP} is exclusively plotted in the Server Time figure as it lacks a distinct Solver phase.

Solver and Server Times scale with increasing fidelity resolution, aligning with the theoretical predictions in Sec.~\ref{subsec:complexity_analysis}. High-resolution operation ($|\mathcal{F}| = 10^3$) yields execution times several orders of magnitude longer than lower settings. This creates a severe scalability bottleneck for the benchmark methods, which require high-resolution to achieve meaningful Ensemble Capacity solutions. At these scales, benchmark Solver Times exceed quantum memory coherence times, rendering them unfeasible for responsive network management. Conversely, CODE bypasses this limitation by extracting near-optimal solutions at a coarser resolution ($|\mathcal{F}| = 10^2$). At this resolution, CODE's Solver Time remains near $10^{-2}$s, keeping Server Time sufficiently low to enable rapid network updates.

\subsubsection{Path Length}
\label{subsubsec:effect_of_path_length} 

\begin{figure}[t!]
    \centering
    \begin{minipage}[t]{\columnwidth}
        \centering
        \includegraphics[width=0.9\columnwidth]{final_figures/plot_legend.pdf} 
        \begin{minipage}[t]{\columnwidth}
        \begin{subfigure}[b]{0.49\columnwidth}
            \centering
            \includegraphics[width=0.95\columnwidth]{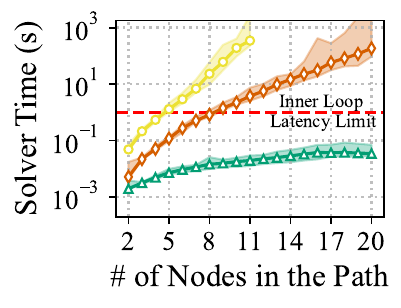}
            \caption{Solver Time}
            \label{fig:solver_time_vs_nodes}
        \end{subfigure}
        \hfill
        \begin{subfigure}[b]{0.49\columnwidth}
            \centering
            \includegraphics[width=0.95\columnwidth]{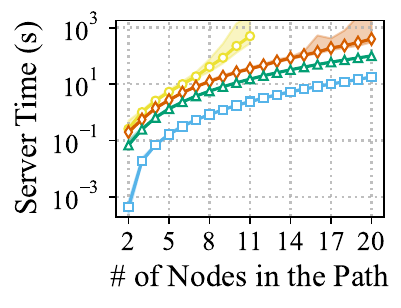}
            \caption{Server Time}
            \label{fig:server_time_vs_nodes}
        \end{subfigure}       
        \end{minipage}
        \vspace{-0.25cm}
        \caption{Running time vs. number of physical links.}
        \label{fig:time_vs_n_nodes}
    \end{minipage}
\vspace{-10pt}
\end{figure}

Next, we evaluate how execution times change as a function of the number of nodes in a path. Using the previous section results, we fix $|\mathcal{F}|=100$ across all solutions and run them for paths lengths from 2 to 20 nodes using the same graph. Figure~\ref{fig:time_vs_n_nodes} presents the 99th, 50th, and 1st percentiles of both Server and Solver Time. As noted previously, \textit{Rate-DP} appears only in the Server Time plot.

From Fig.~\ref{fig:server_time_vs_nodes}, we see that \textit{Rate-DP} exhibits the fastest Server Time among the four solutions. However, this lower computational overhead comes at the expense of a significant performance degradation (see Sec.~\ref{subsec:performance_benchmarking}). \textit{Rate-LP} exhibits the worst server time scalability; its computational cost compounds so aggressively that it mimics exponential scaling before becoming entirely unfeasible for path lengths higher than 11 nodes. CODE shows better Server Times than \textit{Rate-LP} and \textit{EC-LP} solutions. It demonstrates lower Server Times than both, processing 20-node paths in approximately $10^2$ seconds in the worst-case scenario. Since CODE meticulously constructs a pruned hypergraph via DP, its Server Time absorbs an initial overhead that the standard LP approaches do not.

The true impact of CODE's DP hypergraph generation can be clearly visualized in fig.~\ref{fig:solver_time_vs_nodes}. CODE's solver time scales exceptionally well, remaining safely bounded between $10^{-3}$ and $10^{-1}$ seconds. Even at the maximum path length, it operates an order of magnitude below the quantum memory decoherence time and its inner loop latency limit of 1s. As both \textit{Rate-LP} and \textit{EC-LP} use an unpruned hypergraph, they scale worse than CODE. Their Solver Time increases rapidly, breaching the 1s operational latency limit at a path length of $5$ nodes for \textit{Rate-LP} and $8$ nodes for \textit{EC-LP}, \rev{underscoring the DP practical significance}.

\subsubsection{Network Size}
\label{net_size_comp}

\begin{figure}[t!]
    \centering
        \begin{minipage}[t]{\columnwidth}
        \begin{subfigure}[b]{0.48\columnwidth}
            \centering
            \includegraphics[width=0.87\columnwidth]{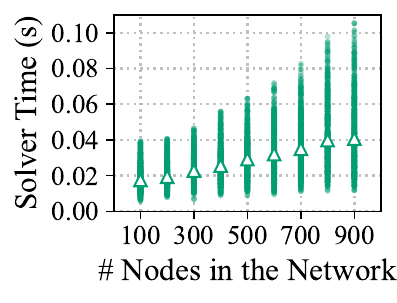}
            \caption{Solver Time}
            \label{fig:entanglement_routing_scalability_1}
        \end{subfigure}
        \hfill
        \begin{subfigure}[b]{0.48\columnwidth}
            \centering
            \includegraphics[width=0.87\columnwidth]{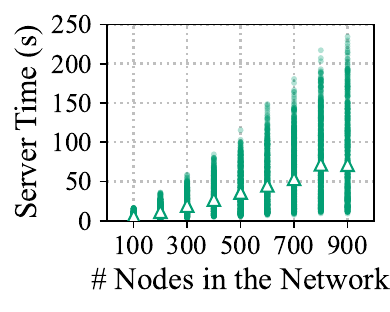}
            \caption{Server Time}
            \label{fig:entanglement_routing_scalability_2} 
        \end{subfigure}
             
        \end{minipage}
        \vspace{-0.25cm}
        \caption{Running time vs. number of nodes in the \gls{qn}.}
        \label{fig:time_vs_nodes_network}
        
\vspace{-3mm}
\end{figure}

Finally, we examine CODE's performance across networks of different network sizes. We conducted extensive experiments using network topologies ranging from 100 to 900 nodes. For each topology, we evaluated 700 randomly selected source-destination pairs, selecting the top 3 shortest paths between them. Fig.~\ref{fig:time_vs_nodes_network} depicts the Server and Solver Times as scatter plots to illustrate the distribution of execution times with a marker in the median performance.

CODE's Server Time ranges from approximately $10^0$ seconds to $10^2$ seconds. Because the Outer Loop is responsible for longer-term, structural network decisions, it is designed to operate on a timescale greater than 1 second. Even at the maximum evaluated network size of 900 nodes, the median Server Time remains computationally reasonable for background processing (peaking around $<10^2$ seconds). This demonstrates that while the state space naturally grows with the underlying infrastructure, CODE's dynamic programming-based generation efficiently manages this expansion.

The operational viability of CODE is validated by its exceptional Solver Time performance in fig.~\ref{fig:entanglement_routing_scalability_1}. Despite the network scaling up to 900 nodes, the LP execution time remains bounded below 0.10 seconds. In fact, the vast majority of instances, including the medians, cluster heavily between $10^{-3}$ and $8\cdot 10^{-2}$ seconds. This remarkable stability highlights the core architectural advantage of CODE: by absorbing the complexity of state space exploration into the non-RT Outer Loop, the Inner Loop is shielded from the combinatorial explosion typically associated with large network topologies. 

%% file: final_docs/9_lim_and_future_work.tex
\section{\rev{Limitations and Future Work}}
\rev{This work relies on an additive lower bound for the Ensemble Capacity and on simplified models of entanglement generation, swapping, and purification; characterizing the gap to the true collective capacity and incorporating more realistic hardware constraints remain open challenges. 
Future work will investigate tighter capacity bounds, collective coding effects, and extensions of the proposed framework to heterogeneous noise, finite-size effects, and experimentally validated network models. 
We also plan to evaluate the scalability and robustness of the optimization framework on larger, dynamic quantum networks and integrate additional hardware and application-level constraints toward practical deployment.}

%% file: final_docs/10_conclusions.tex
\section{Conclusions}
\rev{This work revisits entanglement distribution as an information-delivery problem, optimizing the effective information carried by distributed entanglement rather than treating throughput and fidelity as independent objectives.
Building on this formulation, it introduces CODE, a two-tiered orchestration framework that jointly optimizes entanglement swapping and purification while separating computationally intensive state-space construction from the near-real-time resource-allocation process.
Results demonstrate that this joint optimization yields increasing performance gains as the distribution path grows, with CODE improving the information transmission capabilities up to an $81\%$ for ten-node paths.
Furthermore, CODE substantially reduces execution time through a DP-based offline pruning strategy, keeping the near-RT optimization within the sub-second requirement for networks of at least $900$ nodes. 
Overall, CODE combines an information-centric objective with scalable orchestration improving the quality and timeliness of entanglement distribution.}